\documentclass[journal]{IEEEtran}
\usepackage{tipa}
\usepackage{bbm}
\usepackage{mathrsfs}
\usepackage{cite,url,subfigure,epsfig,graphicx}
\usepackage{amssymb,amsmath}
\usepackage{indentfirst}
\usepackage{algorithmic}
\usepackage{algorithm}
\usepackage{pdfpages}
\usepackage{times,verbatim,amsfonts,amsmath,color}
\usepackage{pifont}
\usepackage{multirow}
\usepackage{booktabs}
\usepackage{adjustbox}
\usepackage[switch,pagewise]{lineno}

\IEEEoverridecommandlockouts
\makeatletter
\def\ps@headings{\def\@oddhead{\mbox{}\scriptsize\rightmark \hfil \thepage}\def\@evenhead{\scriptsize\thepage \hfil \leftmark\mbox{}}\def\@oddfoot{}\def\@evenfoot{}}
\makeatother 

\newcommand{\tabincell}[2]{\begin{tabular}{@{}#1@{}}#2\end{tabular}}

\begin{document}

\title{A Survey on Digital Twins: Architecture, Enabling Technologies, Security and Privacy, and Future Prospects}
%Digital Twin Security: A Survey %A Survey on Internet of Digital Twins: Working Principles, Security and Privacy, and Future Prospects %Principles
\author{
\IEEEauthorblockN{Yuntao~Wang\IEEEauthorrefmark{2}, Zhou~Su\IEEEauthorrefmark{2}\IEEEauthorrefmark{1}, Shaolong~Guo\IEEEauthorrefmark{2}, Minghui Dai\IEEEauthorrefmark{3}, Tom H. Luan\IEEEauthorrefmark{2}, and Yiliang~Liu\IEEEauthorrefmark{2}}\\
\IEEEauthorblockA{
\IEEEauthorrefmark{2}School of Cyber Science and Engineering, Xi'an Jiaotong University, Xi'an, China\\
\IEEEauthorrefmark{3}State Key Laboratory of Internet of Things for Smart City, University of Macau, Macau, China\\
\IEEEauthorrefmark{1}Corresponding Author: zhousu@ieee.org
}}

\maketitle

\begin{abstract}
By interacting, synchronizing, and cooperating with its physical counterpart in real time, digital twin is promised to promote an intelligent, predictive, and optimized modern city. Via interconnecting massive physical entities and their virtual twins with inter-twin and intra-twin communications, the Internet of digital twins (IoDT) enables free data exchange, dynamic mission cooperation, and efficient information aggregation for composite insights across vast physical/virtual entities. However, as IoDT incorporates various cutting-edge technologies to spawn the new ecology, severe known/unknown security flaws and privacy invasions of IoDT hinders its wide deployment. Besides, the intrinsic characteristics of IoDT such as \emph{decentralized structure}, \emph{information-centric routing} and \emph{semantic communications} entail critical challenges for security service provisioning in IoDT.
To this end, this paper presents an in-depth review of the IoDT with respect to system architecture, enabling technologies, and security/privacy issues.
Specifically, we first explore a novel distributed IoDT architecture with cyber-physical interactions and discuss its key characteristics and communication modes.
Afterward, we investigate the taxonomy of security and privacy threats in IoDT, discuss the key research challenges, and review the state-of-the-art defense approaches.
Finally, we point out the new trends and open research directions related to IoDT.
\end{abstract}

\begin{IEEEkeywords}
Internet of digital twins, security, privacy, artificial intelligence, semantic communication, and blockchain.
\end{IEEEkeywords}

\IEEEpeerreviewmaketitle
%----------------------------------------------------------------------------------

\section{Introduction}
Digital twin or cyber twin, as an enabling technology to build future smart cities and the industrial metaverse, has recently spawn increasing global interests from industry and academia \cite{9899718,9120192,8901113}. %popularity
A digital twin means a virtual representation of a real-world entity, system, process, or other abstraction, which can be instanced by a computer program or encapsulated software model that interacts and synchronizes with its physical counterpart \cite{8901113}. With the assistance of digital twins, a variety of intelligent services such as preventive maintenance \cite{9247401}, car accident avoidance \cite{9392784}, ramp merging \cite{9502522}, intelligent maritime transportation \cite{9626558}, and COVID-19 pandemic mitigation \cite{9857637} can be enabled.
Due to its promising future, many tech giants including Meta and Nvidia have declared their ventures into the era of digital twin.
As anticipated by Research\&Markets \cite{DTmarket}, the global digital twin market will reach \$73.5 billion by 2027, with a 60.6\% compound annual growth rate during 2022-2027.

\begin{figure}[!t]\setlength{\abovecaptionskip}{-0.0cm}\vspace{-1mm}
\centering
\includegraphics[width=8.8cm]{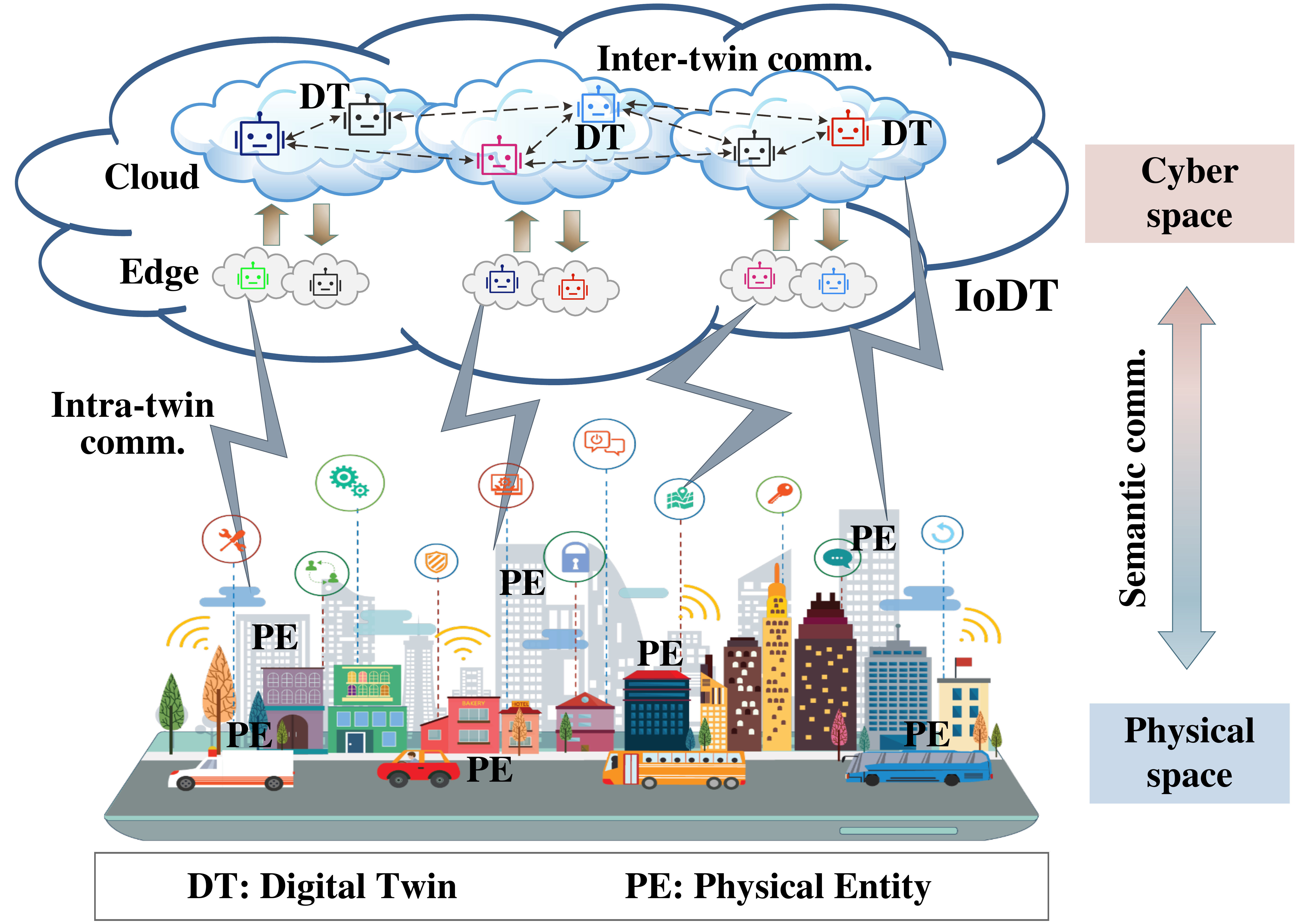}
  \caption{An overview of the Internet of digital twins (IoDT). Digital twin synchronizes with its physical entity via intra-twin semantic communications. Digital twins on cloud/edge servers communicate with each other to share information and knowledge via inter-twin semantic communications. The IoDT connects PEs using the relay of digital twin (DT) communications.}\label{fig:overview}\vspace{-3.5mm}
\end{figure}

With the proliferation of the Internet of things (IoT) infrastructures, billions of things can be represented as digital twins. Then, massive data from connected digital twins can be aggregated to derive composite insights across a vast number of physical entities (e.g., a vehicle, a charging station, or even a city) with dynamic attributes. Eventually, in such shared virtual worlds, users and physical objects are brought together to communicate, interact, and collaborate with digital twins, giving birth to the Internet of digital twins (IoDT). The IoDT is an information sharing network with massive connected physical entities and their virtual twins \cite{luan2021paradigm,9839640,9854866}.
As shown in Fig.~\ref{fig:overview}, in IoDT, physical entities and digital twins can freely exchange information, dynamically synchronize statuses, and cooperatively perform missions with each other through intra/inter-twin communications. For instance, a digital twin city of Shanghai with 26 million inhabitants has been built in 2020 for planning and reacting the COVID-19 pandemic \cite{ShanghaiDT}.

The IoDT incorporates a range of cutting-edge technologies as its foundation. Particularly, artificial intelligence (AI) enables high fidelity and consciousness in mirroring the physical entities and systems; semantic communications provide ultra-low latency semantic transmissions for both intra-twin and inter-twin communications \cite{9955312}; cloud-edge computing and space-air-ground integrated networking (SAGIN) provision massive feasible computing power and ubiquitous networking capacities \cite{yin2022cybertwin}; and blockchain ledgers enforce trust establishment in data/value exchange among virtual/physical twins via decentralized ledgers, distributed consensus, and trust-free smart contracts.

\subsection{Challenges for Securing Internet of Digital Twins}\label{sec:Challenges}
Despite the promising prospects of IoDT, security and privacy concerns pose huge challenges for its wide development.
In IoDT, various security vulnerabilities and privacy breaches may arise from the pervasive individual data collection, massive digital twin data sharing, to the safety of critical infrastructures.
Firstly, digital twin data is usually delay-sensitive and mission-critical. In IoDT, digital twin-related data should travel across multiple networks, softwares, and applications in its lifetime for service offering, making the all-the-round security provision and full-process trust establishment become a challenging issue.
Secondly, to maintain a digital clone of the physical objects, humans, systems and other entities, the personal data to be collected via pervasive IoT devices in the IoDT can be at an unprecedented granularity level and high synchronization frequency, which opens new opportunities for crimes and misuses of private digital twin-related data.
Thirdly, as IoDT is built upon various emerging technologies for service offering, all their security threats and flaws (e.g., eavesdropping, botnets, fraud and phishing) can be inherited by the IoDT.
Lastly, with the growing diversity and complexity in terms of functionalities, brand new and unexpected threats such as semantic data/knowledge poisoning and virtuality-reality synthesized threats can breed in the new IoDT ecosystem.

Due to the intrinsic characteristics of IoDT in terms of \emph{autonomous intelligence}, \emph{decentralized structure}, \emph{information-centric routing}, and \emph{semantic communications}, the security and privacy issues cannot be solely resolved by conventional approaches with the following reasons. %solutions defending against the threats have not kept pace
1) Driven by the interweaving effects of several technologies and the new characteristics of IoDT, the influence of existing vulnerabilities and threats in these technologies can be strengthened and become more severe in IoDT.
2) As digital twin-related services and applications are generally delay-sensitive and mission-critical, it necessitates a tradeoff among service latency, system overhead, and security provision for various IoDT applications with various quality-of-service (QoS) requirements. For instance, how to manage the massive heterogeneous physical entities and their digital counterparts efficiently in IoDT under the decentralized structure remains a challenge.
3) Essentially, IoDT is an extended form of cyber-physical systems (CPS). As the IoDT connects the cyber and physical spaces and remains frequent data synchronization, exchange, and feedback between them, hackers could infiltrate and endanger vital physical infrastructures like power grids and water supply systems by taking advantage of cybersecurity vulnerabilities.
4) The IoDT may raise opportunities for new types of crimes with more covert, hard-to-trace, and cyber-physical synthesized features, which raises huge regulation demands for new laws and regulations in IoDT. For instance, the in-network caching and semantic communication features of IoDT can bring new security threats such as cache pollution, interest flooding, semantic knowledge poisoning, and more implicit privacy disclosures.

\subsection{Comparison with Existing Survey Works and Contributions of Our Survey}\label{subsec:Contributions}
Various research efforts have focused on the promising digital twin. There have been several surveys of the digital twin from different perspectives until now.
For instance, Barricelli \emph{et al}. \cite{8901113} discuss the key concepts, characteristics, and use cases of digital twins.
Fuller \emph{et al}. \cite{9103025} investigate the applications, challenges and existing approaches in applying the digital twin technology into manufacturing, healthcare, and smart cities.
Mihai \emph{et al}. \cite{9899718} comprehensively survey the key enablers, critical challenges, and potential applications of digital twins.
Minerva \emph{et al}. \cite{9120192} systematically review the architectural models as well as the use cases of digital twins in IoT scenarios.
Kuruvatti \emph{et al}. \cite{9923927} survey the potentials and challenges in applying digital twin technology toward constructing future 6G communication systems.
Wen \emph{et al}. \cite{9801816} review existing approaches in realizing digital twins for efficient system and dynamics modeling of the complex networked systems.
Tang \emph{et al}. \cite{9854182} discuss the supporting technologies and key issues in the deployment and update of cyber twins under edge environments.
Alcaraz \emph{et al}. \cite{9765576} investigate four functional layers for digital twin from the data perspective and discuss the security and privacy issues of digital twins in data acquisition, data synchronization, data modeling, and data visualization.
Wu \emph{et al}. \cite{9429703} present the digital twin network, which leverages the digital twin technology to stimulate and predict network dynamics, as well as evolve and optimize network management. Besides, the authors offer an in-depth review of the digital twin network including the key features, technical challenges, and potential applications.
By integrating the emerging digital twin technology and wireless systems, Khan \emph{et al}. \cite{9854866} present a thorough taxonomy including \emph{twins for wireless} and \emph{wireless for twins}.
%,
In contrast to the aforementioned existing survey on digital twins, this survey's goal is to thoroughly discuss the fundamentals, security, and privacy of IoDT including IoDT architecture, key enablers, security/privacy threats, key challenges, and state-of-the-art defenses. A comparison of contributions made by our survey and previous survey works in the field of digital twins is provided in Table~\ref{contribution}.

This paper offers an in-depth review on the system architecture, supporting technologies, security/privacy issues, state-of-the-art solutions, and future trends of the IoDT (i.e., a network of interconnected virtual twins and their physical counterparts along with their attributes and values). Two communication modes, i.e., \emph{inter-twin} and \emph{intra-twin} communications, are presented as well as the security/privacy issues and challenges brought by them during inter-twin, intra-twin, and cyber-physical interactions. The main contributions of this work are three-fold.
\begin{itemize}
  \item We investigate the general architecture, communication modes (i.e., \emph{inter-twin} and \emph{intra-twin} communications), key characteristics (i.e., \emph{autonomous intelligence}, \emph{decentralized structure}, \emph{information-centric routing}, and \emph{semantic communications}), enabling technologies, and modern prototypes of IoDT.

  \item We comprehensively survey the security and privacy threats in the IoDT from seven perspectives (i.e., data, authentication, communication, privacy, trust, monetization, and cyber-physical) as well as the key challenges to resolve them. Besides, the existing/potential security and privacy countermeasures are examined and their feasibilities in IoDT are discussed.
  \item We discuss open research issues and point out future research directions toward building the most efficient and secure IoDT paradigm to enable diverse intelligent applications.
\end{itemize}

\subsection{Organization of Our Survey}\label{subsec:organization}
The remainder of this paper is organized as below. We first offer an overview of the IoDT in Section~\ref{sec:OVERVIEW}. Section~\ref{sec:Threat} and Section~\ref{sec:Defense} discuss the taxonomy of security and privacy issues in IoDT and state-of-the-art security and privacy countermeasures from seven aspects, respectively.
We then outline future research directions in Section~\ref{sec:FUTUREWORK}. Finally, conclusions are drawn in Section~\ref{sec:CONSLUSION}. Fig.~\ref{fig:organization} depicts the organization structure of this survey.

\begin{table}[!t]
   \centering \setlength{\abovecaptionskip}{0cm}
    \caption{A Comparison of Our Work with Relevant Surveys}\label{contribution}
    \resizebox{1.01\linewidth}{!}{
        \begin{tabular}{|c|c|l|}
        \hline
        \textbf{Year.} &\textbf{Refs.} &\textbf{Contribution} \\ \hline %$\dagger$
        {2019} &\cite{8901113} &\tabincell{l}{Discussions on key concepts, characteristics, and use cases \\of digital twins.} \\ \hline

        {2020} &\cite{9103025} &\tabincell{l}{Study on applications, challenges, and existing approaches in \\applying digital twins.} \\ \hline

        {2020} &\cite{9120192} &\tabincell{l}{Review on architectural models and use cases of digital twin \\in IoT applications.} \\ \hline

        {2021} &\cite{9429703} &\tabincell{l}{An in-depth review on digital twin network including key \\features, technical challenges, and potential applications.} \\ \hline

        {2022} &\cite{9899718} &\tabincell{l}{Overview of key enablers, critical challenges, and potential \\applications of digital twins.} \\ \hline

        {2022} &\cite{9923927} &\tabincell{l}{Survey on the potentials and challenges in applying digital \\twins in constructing 6G.} \\ \hline

        {2022} &\cite{9801816} &\tabincell{l}{Survey on digital twins for modeling of complex networked \\systems.} \\ \hline

        {2022} &\cite{9854182} &\tabincell{l}{Discussions on supporting technologies and key issues in \\deploying and updating digital twins in edge.} \\ \hline

        {2022} &\cite{9765576} &\tabincell{l}{Discuss security and privacy issues of digital twins in four\\ functional layers from the data perspective.} \\ \hline

        {2022} &\cite{9854866} &\tabincell{l}{A comprehensive taxonomy in integrating the emerging \\digital twin technology and wireless systems.} \\ \hline

        {Now} &\textbf{Ours} &\tabincell{l}{Comprehensive survey of the general architecture and key \\characteristics of IoDT, discussions on the security/privacy \\threats, critical research challenges, state-of-the-art defenses, \\and open directions in IoDT.} \\ \hline
        \end{tabular}}
\end{table}

\section{Internet of Digital Twins: Working Principles}\label{sec:OVERVIEW}
In this section, we present the general architecture, communication modes, key characteristics, and enabling technologies of the IoDT.

\begin{figure}[h]%!t
\centering \setlength{\abovecaptionskip}{-0.1cm}
  \includegraphics[width=8.5cm]{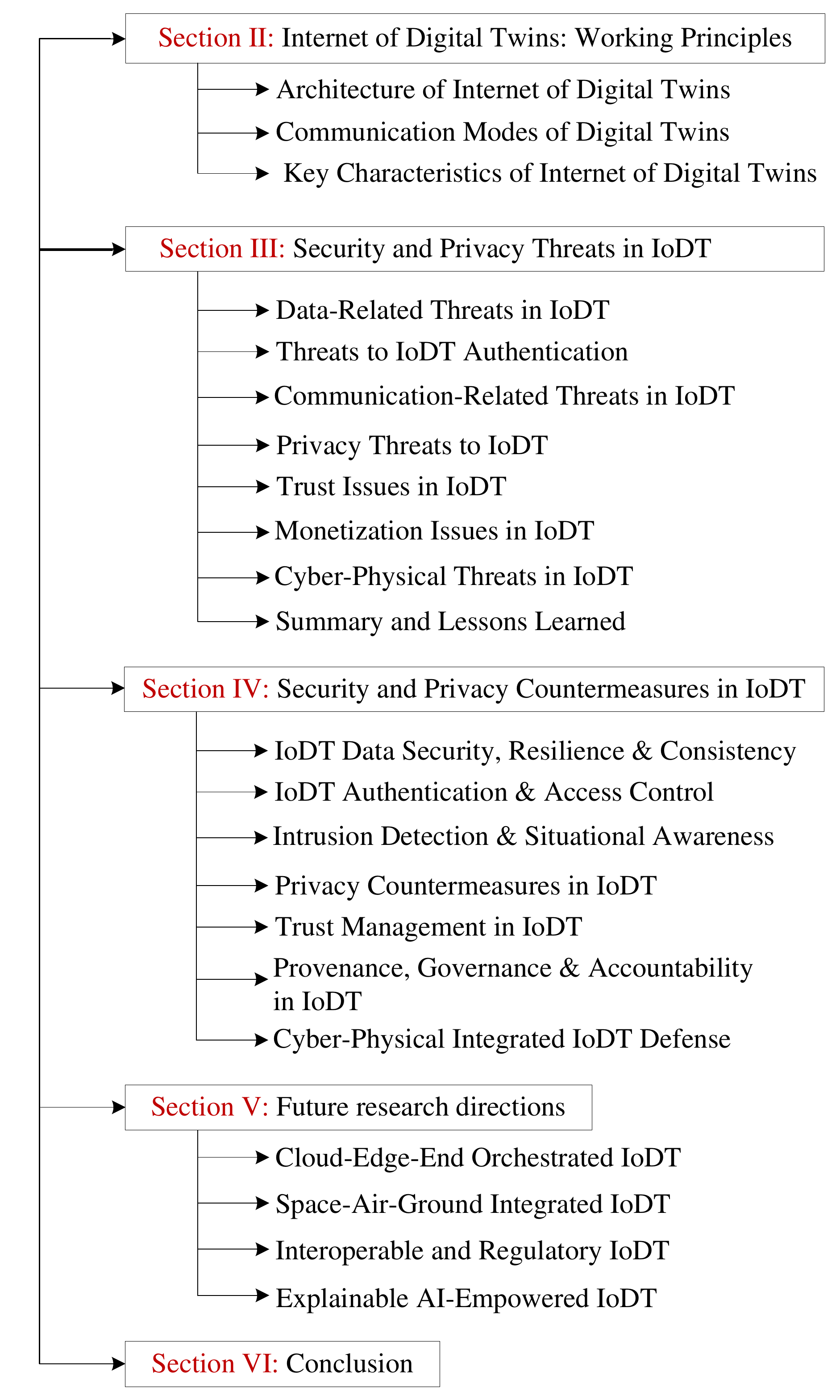}
  \caption{Organization structure of this survey.}\label{fig:organization}\vspace{-2.mm}
\end{figure}

\subsection{Architecture of Internet of Digital Twins}\label{subsec:Architecture}
\begin{figure*}[!t]\setlength{\abovecaptionskip}{-0.0cm}
\centering
  \includegraphics[width=13.5cm]{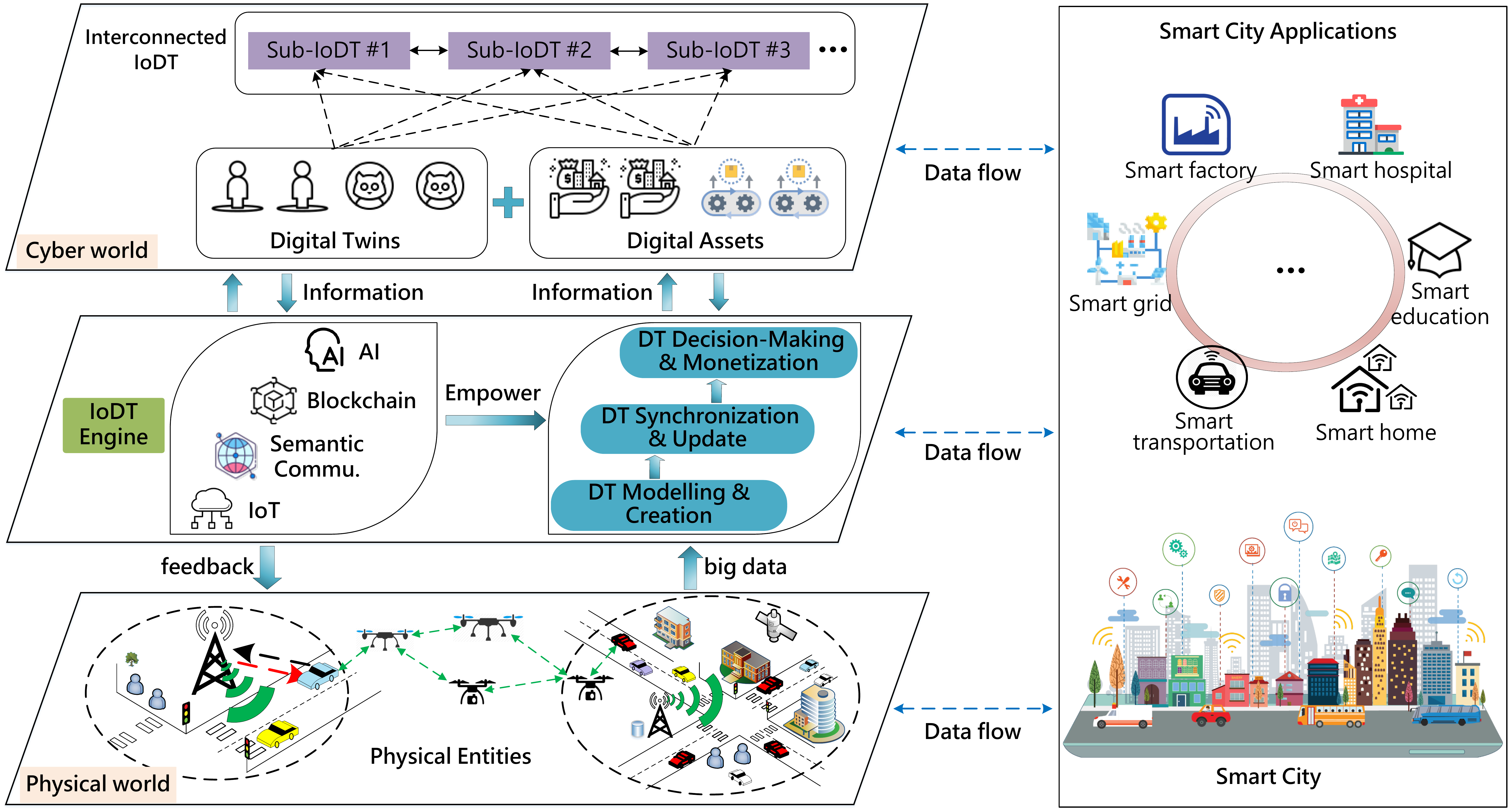}
  \caption{The general architecture of the IoDT in connecting the physical and cyber spaces to empower smart city applications.}\label{fig:architecture}\vspace{-2mm}
\end{figure*}

As shown in Fig.~\ref{fig:architecture}, the construction of IoDT involves the following three elements: (i) the physical entities (PEs) in the real space, (ii) the digital twins along with their virtual assets in the software form in the cyber space, (iii) and an IoDT engine that links the cyber and physical worlds together via the input big data and output feedback.

\emph{Physical Entity (PE)}:
In the physical space, the pervasive PEs can be classified into four main types: sensing PEs, control PEs, hybrid PEs, and infrastructure PEs. Specifically, sensing PEs (e.g., IoT sensors, smart meters, and wearable devices) are obligated for real-time data gathering from things and the environment. For instance, an autonomous vehicle (i.e., PE) can mount multiple advanced sensors including cameras for $360^\mathrm{o}$ environment view and LiDAR for real-time object detection and distance measurement. Control PEs refer to the actuators which execute relevant instructions or actions according to decisions fed back from the cyber layer. Hybrid PEs are the ones who serve as both roles concurrently.
Infrastructure PEs contain the grid infrastructures, networking infrastructures, computing infrastructures, etc. Grid infrastructures such as power lines offer urban/rural electricity, networking infrastructures offer wireless/wired communication capacity, while computing infrastructures provide computation, caching, and storage capacities.

\emph{Digital Twin}:
In cyberspace, a virtual representation of the real-world entity, system, process, or other abstraction is known as a digital twin \cite{9899718}. It can be instanced by a computer program or a software model which interacts and synchronizes with its physical counterpart in real time. Besides, the digital twin can be deployed within a cloud or an edge server \cite{luan2021paradigm}.
A synchronized private link can be established for real-time data transmission between the digital twin and its PE or other twins \cite{9852383}. In addition to being able to instantly visualize the status of their PEs, digital twins can also help their physical counterparts make anticipatory operations, thereby enabling intelligent services such as 3D simulation, preventive maintenance, and smart decision-making.
For instance, a digital twin of a vehicle can learn the personalized preferences of the vehicle user, download the interested vehicular media from other twins on the road, and accurately plan the driving trajectory based on the synchronized vehicular information (e.g., speed, direction, and surroundings), regional traffic information, and weather conditions.

\emph{Internet of Digital Twins (IoDT)}:
As shown in Fig.~\ref{fig:architecture}, the IoDT is generally composed of multiple interconnected sub-IoDTs. In the IoDT, billions of connected virtual twins can freely share information, dynamically synchronize statuses with physical objects, and cooperatively perform missions with each other, thereby forming an information sharing network with numerous potentials. In such shared IoDT, massive distributed data shared by various digital twins can be effectively aggregated to obtain composite insights across a vast number of physical entities (e.g., a vehicle, a charging station, or even a city). Additionally, with the help of digital twins and the IoDT, users and physical objects can be brought together to communicate, interactive, and collaborate with digital twins.
For instance, for two physical vehicles that tend to learn the road traffic from each other, when their direct vehicle-to-vehicle (V2V) connections are unavailable due to the out-of-field, their digital representatives can freely communicate and interact with each other to enable more efficient data exchange.

\emph{IoDT Engine}:
Because of the bidirectional connection between PEs and their digital twins, the IoDT engine feeds the PEs' private data to model, create, maintain, and update the digital representatives along with the virtual assets. The IoDT engine is created through the convergence of various emerging technologies including IoT, AI, semantic communication, and blockchain.
\begin{itemize}
  \item \emph{IoT}. The IoT is built on a combination of several technologies, including general-purpose computing, commodity sensors, machine learning, and increasingly powerful embedded systems. IoT is the underlying technology of IoDT, which offers the sensing/networking/computing infrastructures and capacities to PEs. The pervasive IoT sensors carry out real-time data collection from things and the environment to the IoDT engine. The cloud-edge computing paradigm provisions massive feasible computing power to enable massive data analysis, data storage, and modeling \cite{9664267}. The SAGIN paradigm offers ubiquitous networking capacities for seamless data exchange/transmission within IoDT \cite{yin2022cybertwin}. A digital twin can associate with multiple physical IoT devices. For example, the twin of an autonomous vehicle can be created and updated by efficiently fusing the multi-source and multi-modal data from multiple advanced sensors such as cameras, radars, and LiDAR.
  \item \emph{AI}. By learning from historical and real time data, AI algorithms enable high-accuracy and real-time simulations to produce and evolve digital twins with high fidelity and consistency in mirroring the physical entities, processes, and systems. For instance, AI models can help predictive maintenance and accident traceability, thereby improving efficiency and reducing risks for industry applications. For efficient multi-twin cooperation in task completion, transfer learning techniques allow twins to use the knowledge learned from other twins (i.e., source domain) to help its learning tasks in the target domain. Through efficient knowledge/parameter sharing between multiple tasks performed by different twins, multi-task learning allows twins to learn multiple correlated tasks simultaneously to enhance the performance and generalization of the trained model on each task. Meta-learning (or learning-to-learn) \cite{9428530} enables twins to learn from the output of other AI algorithms which learn from historical data/experience, thereby making a prediction given predictions made by other AI algorithms. By incorporating deep learning and reinforcement learning (RL), deep RL (DRL) allows twins to make optimal decisions from unstructured input data in complex and dynamic environments via trails. Moreover, multi-agent RL (MARL)\cite{4445757} enables various twins (whose PEs coexist in a shared environment) to make individually optimal decisions with multi-agent effects, where each twin is motivated by its own rewards to advance its own interests.
  Besides, distributed AI technologies such as federated learning \cite{9928220} allow efficient data aggregation and sharing across various digital twins to derive insightful results.
  \item \emph{Semantic communication}. In IoDT, there exist massively frequent data synchronization interactions between PEs and digital twins, as well as the intensive data exchanges between twins, raising huge demands for low-latency and low-overhead communications. Semantic communication \cite{9252948,9830752}, as the breakthrough beyond the Shannon paradigm, provides a promising solution by offering ultra-low latency semantic transmissions for both intra-twin and inter-twin communications, where only the meaningful data essential for the task are transmitted.
  \item \emph{Blockchain}. The blockchain technology \cite{9631953} offers decentralized ledgers, distributed consensus protocols, and trust-free smart contracts to automatically enforce asset identification and ownership provenance as well as trust establishment in data/value exchange among virtual twins. Via hash-chained blocks and sophisticated cryptography, the stored data in historical blocks can be immutable and irreplaceable, ensuring the data/record reliability. The non-fungible token (NFT) empowered by blockchain ledgers can determine authentic rights (e.g., asset identification and ownership provenance) for virtual assets in the IoDT market and help construct the economy system in IoDT. The distributed consensus protocols can help IoDT governance and regulation in a democratic and efficient fashion. Besides, the smart contracts allow automatic and trust-free exchange of data, knowledge, resource, and asset among virtual twins.
\end{itemize}

The IoDT engine can be solely or collaboratively deployed at the digital twin side, PE side, and networking/computing infrastructure side, depending on specific digital twin applications.
Informally, in the IoDT, AI serves as the ``brain'', IoT is the ``bone'', semantic communication acts as the ``ears'', and blockchain is the ``blood'', thus connecting the whole digital twin ecosystem.

\begin{figure}[!t]\setlength{\abovecaptionskip}{-0.0cm}%\vspace{-3mm}
\centering
\includegraphics[width=9.8cm]{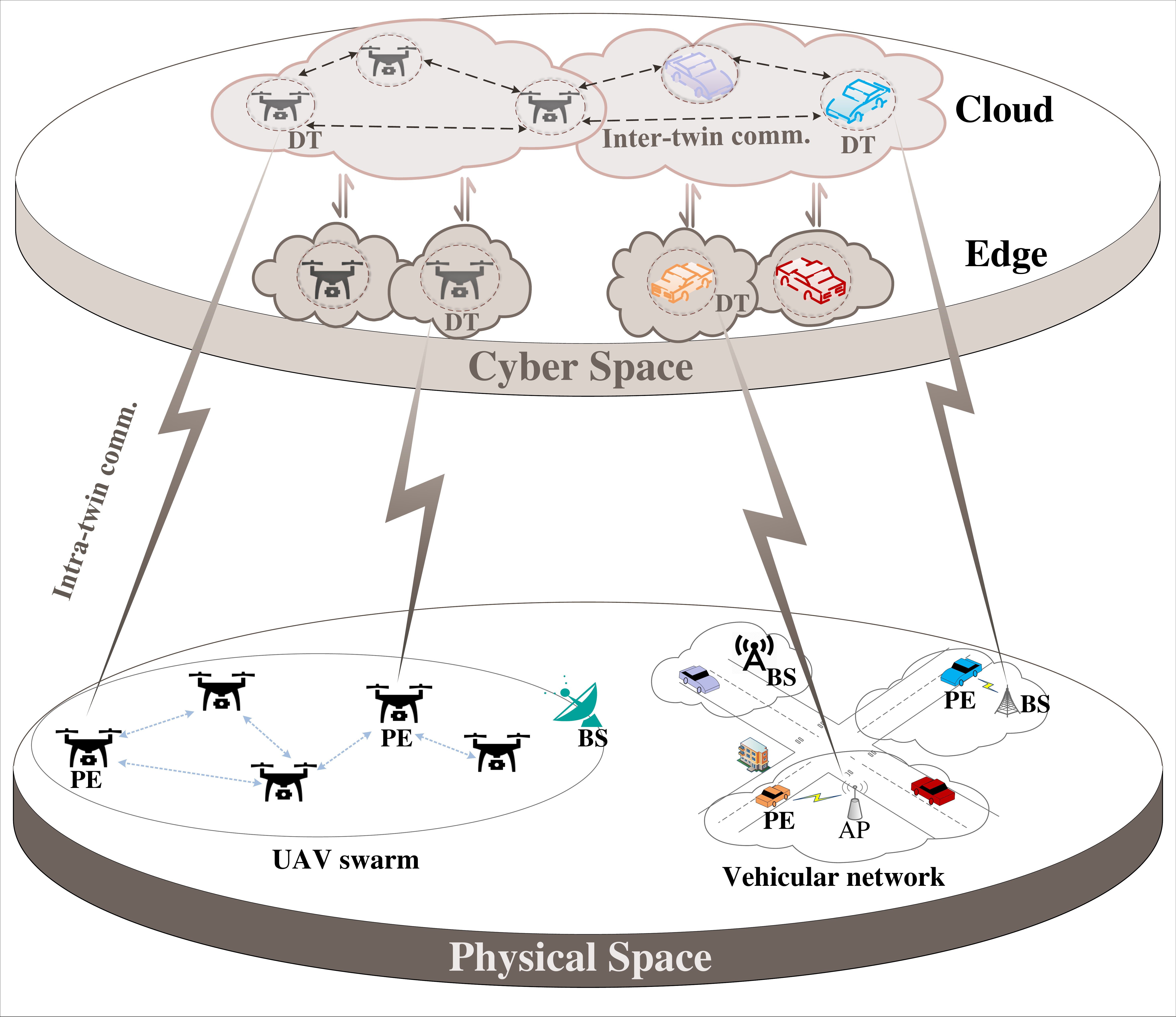}
  \caption{Illustration of inter-twin communication and intra-twin communication in the IoDT.}\label{fig:TwinComm}\vspace{-2mm}
\end{figure}

\subsection{Communication Modes of Digital Twins}\label{subsec:CommMode}
In the IoDT, as shown in Fig.~\ref{fig:TwinComm}, there exist two types of communication modes \cite{luan2021paradigm,9852383}, i.e., \emph{inter-twin communication} for data synchronization between PEs and twins and \emph{intra-twin communication} for coordination and cooperation between twins.

\begin{itemize}
  \item \emph{Inter-Twin Communication}: Digital twins in the cyber space can spontaneously discover and obtain necessary information from other twins based on the PE's requirements. A inter-twin connection can be established for data access and data sharing activities between two twins. As twins are located in the cloud/edge environment, the inter-twin communication thereby breaks the space-time limits in the real space and facilitates data transmission and collaboration activities for PEs that are originally located far away.
  \item \emph{Intra-Twin Communication}: The intra-twin communication bridges the PE and its digital twin, by building private data flow links between them. Essentially, virtual twins are driven by the PEs' real-time raw data; moreover, PEs are optimized by the feedback and smart decisions of digital twins. For instance, in IEEE 1451 smart sensor digital twin federation \cite{8706111}, the digital twin of a real-world IEEE 1451 smart sensor can intelligently simulate the behaviors and failure modes of its PE via intra-twin data communication. Intra-twin communication is featured with bidirectionality with different synchronization levels. Bidirectionality refers to two-way interactions between PE and its virtual twin. Besides, different services can have versatile synchronization requirements ranging from real-time ($\thicksim$millisecond) to near real-time ($\thicksim$second) and to delay-tolerant ($\thicksim$minute).
\end{itemize}

\emph{Illustrating Example.} As shown in Fig.~\ref{fig:TwinComm}, there are multiple unmanned aerial vehicles (UAVs) and ground vehicles involved in a common traffic scheduling task based on IoDT. Considering the unpredictable dynamics of aerial UAVs and ground vehicles and the dynamic communication connections between UAVs and vehicles, it is challenging to monitor the real-time on-road traffic for efficient traffic scheduling and path planning. Instead, digital twin UAVs in the cloud can efficiently obtain traffic information from other twin UAVs and twin vehicles via inter-twin communications, thereby breaking the limitations of physical communication range and intermittent aerial-ground links. Moreover, based on the task-relevant information and continuous semantic data flow from its physical counterpart, the virtual twin UAV can dynamically learn and predict the location of its PE and autonomously make decisions on the related sensors (e.g., angle of camera) on its PE to help complete the traffic scheduling mission.

\subsection{Key Characteristics of Internet of Digital Twins}\label{subsec:Characteristics}
The IoDT exhibits the following key characteristics to construct a flexible information sharing system for diverse smart applications.

\subsubsection{Autonomous Intelligence}
In the IoDT, digital twins can proactively seek the valuable information from relevant twin nodes via inter-twin connections for intelligent decision-making without notifying their PEs. Moreover, after being granted, digital twins can autonomously connect to their PEs for real-time synchronization without being instructed. Essentially, given sufficient data and computing power supply, digital twins can work autonomously as intended.

\subsubsection{Decentralized Structure}
As digital twins are virtual and autonomous agents, the data transmissions between twins are spontaneously provoked without being instructed by the central manager. Moreover, there exist no central server for the management of massive heterogeneous twin nodes. Besides, the data transmissions between twins are generally delay-sensitive, where the centralized networking paradigm may lead to unnecessary data hops and extra data latency.
Hence, the data exchange between digital twins are executed in a peer-to-peer (P2P) cooperative manner in the IoDT. Additionally, the feedback produced by digital twins can be forwarded to the corresponding PE via intra-twin connections.

\subsubsection{Information-Centric Routing}
In the IoDT, digital twins are more concerned about how to fast retrieve useful information from relevant twin nodes, instead of from which specific data source for data retrieval. Compared with current IP-based host-oriented Internet, the information-centric routing mode (e.g., publish/subscribe (pub/sub) paradigm \cite{9397267} and named data networking (NDN) \cite{8027034}) can benefit digital twins to rapidly retrieve the demanded information in the large-scale IoDT based on the interests, via uniquely named data and in-network caching. Data in IoDT is independent of its source, application, and means of transmission and can be directly addressable and routable, thereby supporting in-network replication and multicast traffic. The digital twin can issue an interest message for content request, and the twin that caches the demanded contents will reply and return them to multiple requesters, which significantly facilitates data exchange between digital twins with reduced content retrieval latency and network loads.
\begin{itemize}
\item \emph{NDN}. In the NDN paradigm, hierarchical naming is widely adopted, and an interest packet can be sent to the IoDT by a user to call for the desired content by its naming information \cite{8027034}.
A NDN router maintains a content store (CS), a pending interest table (PIT), and a forwarding information base (FIB) \cite{8027034}.
Once the forwarding router receives the interest, it searches for its CS using the content name and returns the requested content if the CS match is successful.
Whenever the desired content is unavailable in its CS, the router checks its PIT to see if there are any previous entries for the content request.
If PIT matches successfully, the interest entry is added to its PIT. If there is no PIT match, a new PIT entry of this interest will be created and this interest will be forwarded. Finally, the content returns to its requester via the interest's inverse path.
\item \emph{Pub/Sub}. In the pub/sub paradigm, the flat naming is widely adopted, which includes a topic ID and a unique content ID \cite{9397267}. A publisher can advertise its content by sending its local broker a \emph{Publish} message, and the broker will route the message to the designation broker who will store the content. A subscriber who is interested in the content object can send its local broker a \emph{Subscribe} message, and this message will be routed to the designation broker. The routing decision of the local broker can be made via a distributed hash table (DHT) \cite{9397267}. Between the publisher and the subscriber, a content delivery path is produced by the topology manager via routing Bloom filters to complete content delivery through intermediate forwarders.
\end{itemize}

\subsubsection{Semantic Communications}
\begin{figure*}[!t]\setlength{\abovecaptionskip}{-0.0cm}%\vspace{-3mm}
\centering
  \includegraphics[width=12.7cm]{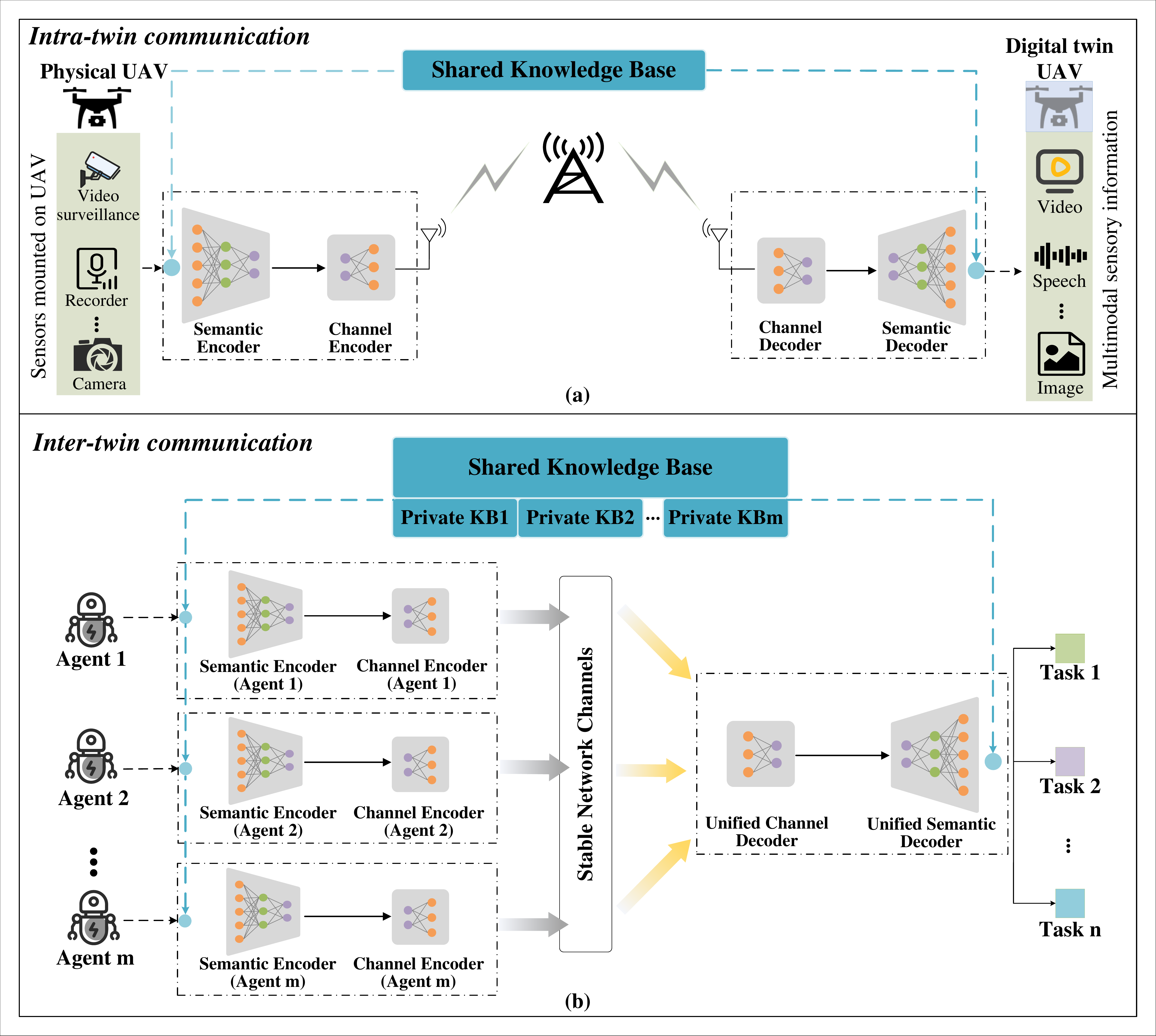}
  \caption{Illustration of semantic communications for intra/inter-twin communications in the IoDT. (a) Intra-twin communication: end-to-end semantic communication between the digital twin and the physical entity, which includes multi-tasking from multiple sensors (e.g. image, video, voice transmission); (b) Inter-twin communication: multi-agent semantic communication among multiple virtual twins in the IoDT.}\label{fig:SemanticComm}
\end{figure*}

Traditional Shannon communication paradigms mainly focus on the accurate transmission of the massive bit sequences. By leveraging AI capacities into communication systems, semantic communications allow transmitting the useful task-relevant information from the source node to the receiver \cite{9830752}, thereby greatly alleviating the data traffic in both inter-twin and intra-twin communications. For instance, in the transmission of a bird picture, rather than transmitting the whole image, the features relevant to recognize the bird (i.e., ``meanings'' of picture) are extracted by a semantic transmitter while irrelevant data (e.g., picture background) is omitted for minimized data transmission without performance degradation. Moreover, using a matched knowledge base (KB) between the sender and the receiver, the sent semantic information can be successfully ``interpreted'' by the receiver \cite{9252948}. Fig.~\ref{fig:SemanticComm} illustrates the intra-twin semantic communications and inter-twin semantic communications in IoDT.
\begin{itemize}
        \item \emph{Intra-Twin Semantic Communication}.
        As illustrated in Fig.~\ref{fig:SemanticComm}(a), intra-twin communication involves data transmission and information interaction between PEs and digital twins. Taking UAV as an example, it has multiple types of sensors, and needs to transmit multi-modal data (e.g., video, speech, and text) \cite{xie2022taskorienteda}. For efficient semantic communication, a prerequisite is that both sending and receiving parties have the same or similar background knowledge \cite{shi2021semantic}; otherwise, communication between users with a high level of knowledge gap (e.g., adults and children) will be inefficient. For intra-twin communication, the same KB is privately shared between the PE and the twin to attain real-time and efficient synchronization. With the help of semantic KB and powerful deep neural networks (DNNs), semantic encoder performs semantic extraction of source information. On the one hand, it can extract task-relevant information and then improve communication efficiency. On the other hand, semantic information irrelevant to the transmission task can be filtered out and compressed, thereby reducing the consumption of communication bandwidth \cite{gunduz2022beyond}. To resist the effects of noise, fading, and interference in the wireless channel, the encoded semantic signal is then passed through a channel encoder to improve the robustness of the system. The encoded signal is transmitted to the receiver over the wireless channel. Guided by the shared KB, the receiver can efficiently reconstruct semantic information from the transmitted signal.
        \item \emph{Inter-Twin Semantic Communication}.
        In IoDT, for a specific intelligent task (e.g., traffic analysis and path planning), the participating twins can cooperate to complete it. In this way, it makes full use of the information possessed by each twin and achieves better semantic reconstruction performance \cite{xie2022taskorienteda}. Specifically, as shown in Fig.~\ref{fig:SemanticComm}(b), the knowledge generally acknowledged and comprehended by multiple agents is stored in the shared KB. Meanwhile, each agent updates its own KB to store the knowledge that is private or shared only with certain agents. Before transmission, each agent performs semantic and channel coding with the aid of the KB, to acquire a semantic representation of the source data which is resistant to channel distortion. Then, the task-relevant semantic information is sent to the server/receiver through a stable network channel. To further exploit the semantic-level correlation of information in the agents (e.g., cameras on different entities capturing images of the same object from different perspective \cite{xu2022semantic}) at the receiver side, a collaborative unified decoding-based module will jointly recover and exploit this semantic information to obtain information for different tasks.
      \end{itemize}

Table~\ref{SemanticCommTable} summarizes the comparison of semantic communications for intra-twin and inter-twin communications in IoDT.

\subsubsection{Heterogeneous Components}
In IoDT, the digital twins are generally Heterogeneous in terms of PE types, software implementations, access interfaces, communication modes, and data types (e.g., provisional and operational). Moreover, there exist different modes in producing digital twins such as on-demand, subscription-based, event-triggered, etc. From the perspective of both hardware and software, the heterogeneous components also contribute to the terrible interoperability of digital twin systems.

\begin{table}[!t]
   \centering
    \caption{A Summary of Semantic Communications For Intra-Twin And Inter-Twin Communications In IoDT}\label{SemanticCommTable}
\begin{tabular}{|l|l|l|}
\hline
                    & \textbf{Intra-twin Comm.} & \textbf{Inter-twin Comm.} \\ \hline
Connection          & One-to-one connection     & Multi-agent connection    \\ \hline
Data Type           & Multimodal                & Multimodal                \\ \hline
Channel             & Wireless channel            & Stable wired channel            \\ \hline
KB & Fully synchronized     & Public \& Private    \\ \hline
\end{tabular}
\end{table}

\section{Security and Privacy Threats in IoDT}\label{sec:Threat}
This section presents a taxonomy of security/privacy threats in the IoDT from the following perspectives: data, authentication, communication, privacy, trust, monetization, and cyber-physical.

\subsection{Data-Related Threats in IoDT}\label{subsec:threat1}
Data flows are essential to build accurate and up-to-date digital twins, and the data life-cycle in the IoDT includes data collection, storage, service, and management.

\begin{figure*}[!t]\setlength{\abovecaptionskip}{-0.0cm}%\vspace{-3mm}
\centering
\includegraphics[width=11.1cm]{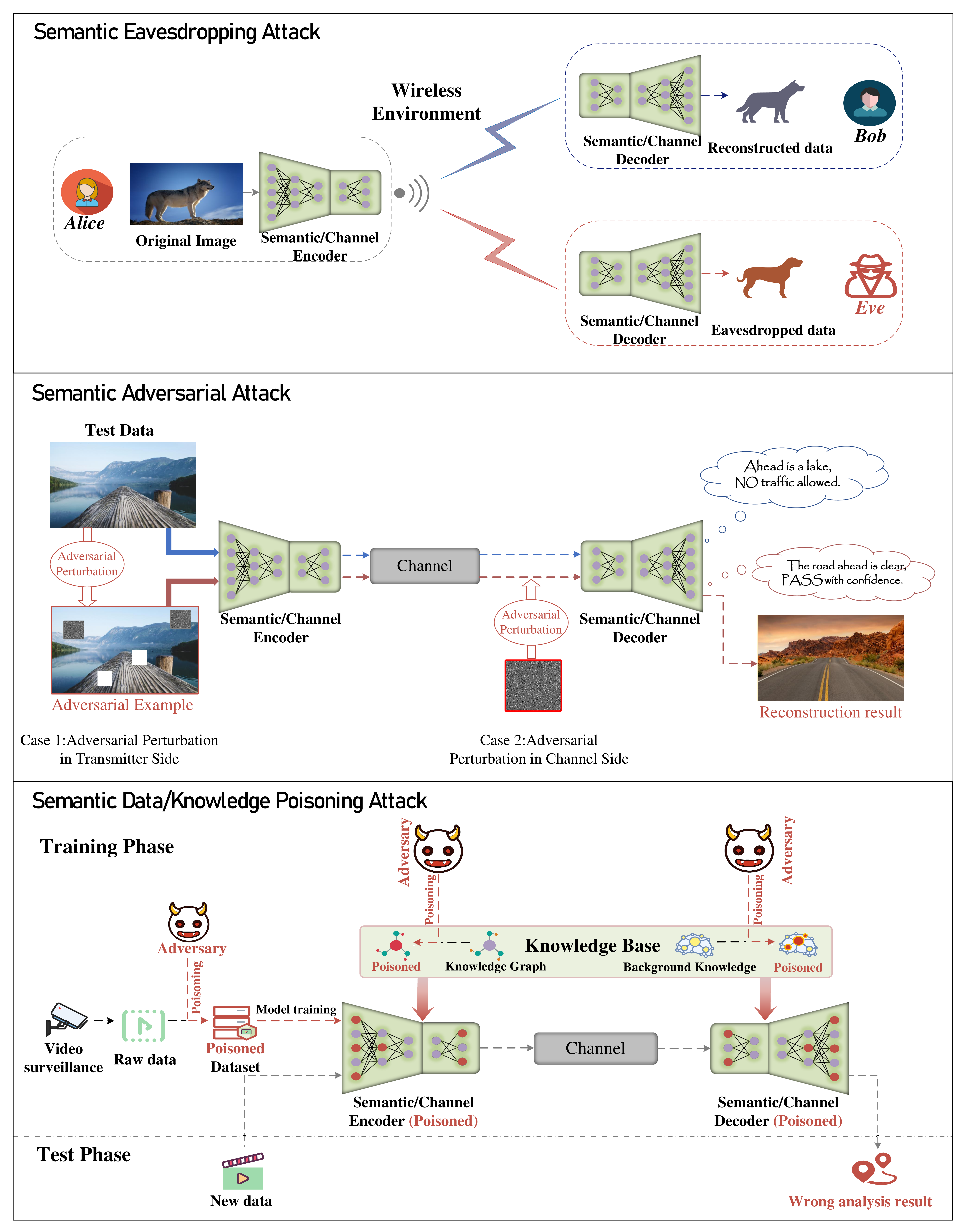}
  \caption{An illustrative example of semantic eavesdropping attack, semantic adversarial attack, and semantic data/knowledge poisoning attack in the IoDT.}\label{fig:attack}\vspace{-2mm}
\end{figure*}

\begin{itemize}
  \item \emph{Data Tampering Attack}. In the life-cycle of digital twin services, the data stream may be forged, modeified, replaced, or removed by attackers in the IoDT. For instance, falsified data can be transmitted to the cyberspace during the digital twin creation process, resulting in erroneous or inconsistent reactions from the digital twins.
  \item \emph{Low-Quality Data Threat}. This attack can occur in both intra/inter-twin interactions. On one hand, the reliability level that a digital twin can mirror and predict its PE depends on the quality of data upon which its simulation models are built, as well as the accuracy and consistency of the models. On the other hand, selfish twins may share low-quality data with other twins in inter-twin cooperation for reduced cost.
  \item \emph{Desynchronization of Digital Twins}. Adversaries may compromise the consistency of digital twins in terms of fidelity and granularity by prioritizing the attack policies and modifying the synchronization frequency in intra-twin interactions \cite{8822494}. For instance, hackers can produce misconfigurations in the monitoring missions to successfully desynchronize the digital twins in the virtual space with respect to the real space. Via the desynchronization of virtual twin models, attackers can disrupt, modify or falsify the constructed digital twins while remaining undetected by removing corresponding log files in the virtual space.
  \item \emph{Model Inconsistency Attack}. A malicious server may distribute different model parameters to different participants (i.e., twins) to manipulate the twin model training process and infer the privacy of twins in inter-twin cooperation \cite{pasquini2022eluding}. For instance, in the personalized digital twin model training process under personalized federated learning, a compromised cloud/edge server may maliciously provide different versions of elaborately designed gradients to participants, which causes the model inconsistency and infers the local gradients of the targeted participant.
  \item \emph{Data/Content Poisoning Attack}. In IoDT, the data/content poisoning attack can be carried out in both data routing and data reasoning processes during inter-twin interactions \cite{9397267}. During data routing in the information-centric IoDT, attackers may fill the CS of a relay node (e.g., access point or edge server) by injecting bogus or worthless contents to the IoDT with valid names for the interests. Moreover, in the data training process, adversaries may alter the distribution of training data, modify the label values (via label contamination), and even inject poisoned or adversary samples, with the aim to produce invalid and erroneous inference.
  \item \emph{Semantic Adversarial Attack}. This attack can occur during both intra/inter-twin interactions. It is also known as semantic test-time evasion attack, which occurs in the inference stage. In conventional human-human communication, adversarial examples have a weak impact on communication accuracy. But for semantic communication between agents, the utility largely depends on the performance of DNNs, which are vulnerable to adversarial examples. As shown in the middle part of Fig. \ref{fig:attack}, there are two ways to implement adversarial attacks during communication. One occurs in the transmitter side \cite{sagduyu2022semantic}, where the adversary affects the subsequent task by adding adversarial perturbations to the raw data. The other is in the channel side. With the integration of computing and communication, computing tasks will be exposed in the open space, which considerably increases the possibility of adding perturbations to the data to become an adversarial example. Semantic adversarial attacks can bring great security risks to IoDT. For instance, an unmanned vehicle detects a lake ahead that is impassable. When DTs construct the virtual environment through the information transmitted by the vehicle, malicious adversaries can mislead DTs into believing the road ahead is clear through adversarial perturbations, resulting in a traffic accident.
  \item \emph{Semantic Data/Knowledge Poisoning Attack}. This attack can occur during both intra/inter-twin interactions. In the semantic communication between twins and PEs or between twins, malicious entities consciously inject poisoned data samples into the raw data or KB, thus serving the purpose of manipulating model training, as depicted in the lower part of Fig. \ref{fig:attack}. Data poisoning usually occurs at the transmitter, where malicious entities utilize contaminated datasets to degrade the performance of DNNs. For instance, a malicious autonomous vehicle may deliberately share erroneous traffic jams to clear the road for itself. Except for channel noise, semantic communication has its own unique semantic noise \cite{hu2022robust}, which creates semantic ambiguity in their understanding of the task. Malicious users can increase semantic noise by injecting specific task-irrelevant knowledge into the KB. For instance, if a PE wants to transmit information about apples (fruit) to the twin, but rich knowledge about digital products is injected into the KB, the twin will probably understand it as apple incorporated.
  \item \emph{Model Poisoning Attack}. In inter-twin interactions in IoDT, adversaries may also modify or replace the immediately trained gradients or AI model parameters via careful calculation to deteriorate the knowledge inference performances of other collaborative learners. For instance, for the digital twin models built on federated learning paradigms, malicious participants may upload Byzantine local AI model updates to mislead the global model aggregation results \cite{tian2023comprehensive}.
  \item \emph{Cache Poisoning/Pollution Attack}. In the information-centric IoDT, to facilitate in-network content caching and replication, each router or host maintains a local cache to lookup and satisfy incoming content requests. A malicious entity may manipulate the local cache of routing nodes (e.g., edge servers and access points) to determine what contents to cache \cite{8887204} in inter-twin interactions. Adversaries may perform cache poisoning and pollution attacks by introducing malicious or unpopular contents/interests into local caches (i.e., cache poisoning) and disrupting cache locality (i.e., cache pollution) \cite{8027034}. The simplest manner to launch cache poisoning/pollution attacks is to vary the popularity distribution of cached contents by frequently requesting non-popular contents, such that non-popular or even invalid contents can be cached in the CS.
  \item \emph{Threats to Data Backup}. Data backup is essential to prevent data losses and corruptions under disasters (e.g., lightning and flood) to enforce data availability and consistency during the life-cycle of digital twin services \cite{8395067}. Adversaries may interfere with or disrupt the backup process to falsify the original digital twin data as intended.
\end{itemize}

\subsection{Threats to IoDT Authentication}\label{subsec:threat2}
\begin{itemize}
  \item \emph{Impersonation Threat}. Adversaries may exploit the system flaws in the authentication phase to impersonate another legitimate identity to extract user's critical information (e.g., credentials or security parameters) in both intra/inter-twin interactions \cite{8678450}.
  \item \emph{Unauthorized Data Access}. This attack occurs in both intra/inter-twin interactions. To empower the intelligent services built on digital twins, various new types of user information (which can be personal and sensitive) are required to be collected in real time and fine granularity \cite{9268472}. After impersonation attack, the malicious users or service providers can gain unauthorized access to the myriad sensitive user information to facilitate targeted ads and precision marketing.
  \item \emph{Unauthorized Knowledge Base Access}. This attack occurs in both intra/inter-twin interactions. For multi-agent communication, there are two types of KBs: one is a public KB accessible to all agents, and the other is an agent-private KB. When a malicious user or service provider unauthorizedly accesses either KB, or even maliciously tampers with its contents, it will greatly affect the performance of semantic communication and leak the user's privacy information.
  \item \emph{Backdoor Attack}. Malicious or disreputable manufacturers may insert compromised components or codes into devices/softwares as backdoors for specific purposes. For instance, they may interrupt the normal operations of the compromised device and cause malfunctions or information leaks.
  \item \emph{Rogue IoDT Devices/Servers}. Rogue devices may maliciously clone and replace the legitimate virtual assets or maliciously update software components of digital twins \cite{golde2012weaponizing}. For rogue servers, as data replicates of massive PEs can be managed by them, they may take control of the digital threads and modify the digital twins to affect the digital space. For instance, rogue gateways, as part of the edge infrastructure, can entail severe privacy leaks and facilitate subsequent threats such as denial of service (DoS).
  \item \emph{Rogue Virtual Assets}. Hackers can insert malicious virtual assets (e.g., containers and virtual machines (VMs)) or replace the legitimate assets with malicious ones with the help of insiders to control a part of the digital twins \cite{9765576}. Then, by exploiting the rogue virtual assets in the virtual space as a springboard, subsequent invasion to control the entire digital twin model, as well as transitive attacks on other digital twin models can be facilitated.
  \item \emph{Privilege Escalation Threat}. Insiders with full rights to access the intranet or external attackers may escalate their privileges by exploiting system flaws (e.g., malware, reverse engineering, and buffer overflows) \cite{monshizadeh2014mace}. Besides, collusive external adversaries can launch attacks such as advanced persistent threats (APT) to invade the insider network and gain illicit access to the target resource. Thereby, highly sensitive user data can be leaked and the main vulnerabilities (e.g., zero-days) in the digital twins of critical infrastructures can be identified.
\end{itemize}

\subsection{Communication-Related Threats in IoDT}\label{subsec:threat3}
\begin{itemize}
  \item \emph{Eavesdropping Attack}. An eavesdropper may eavesdrop open and unsecured communication channels to access the transmitted data such as the semantic information between PEs and twins and between virtual twins.
  \item \emph{Semantic Eavesdropping Attack}. This attack occurs in both intra/inter-twin interactions. In conventional communication systems, it is challenging for eavesdroppers to derive the privacy information from the channel containing a number of noise. Semantic communication can still achieve better performance under low SNR \cite{luo2022encrypted}, but it also brings opportunities for eavesdroppers, as depicted in the upper part of Fig. \ref{fig:attack}. In the case of poor channel conditions, eavesdroppers can still decipher semantic information with the help of a shared decoder. Moreover, semantic information can reflect users' real data distribution to a certain extent, making it simpler to expose user privacy.
  \item \emph{Message Flooding Attack}. During intra/inter-twin cooperations, adversaries may send or forward a large number of flooding messages in the IoDT to cause a DoS. The flooding messages can be comparatively simple, but if there are enough, it can make the twin node severely disabled.
  \item \emph{Interest Flooding Attack}. During the life-cycle of digital twin service, an adversary may send thousands of interest packets (which are not sufficiently resolved or not resolved at all) for content request in information-centric IoDT to cause malicious CPU or memory consumption, thereby overloading the network infrastructure \cite{8027034}. For instance, collusive adversaries may produce multiple interests with random names (flat or hierarchical) to cause the traffic jam of the wireless network, hence denying digital twin services to legitimate users.
  \item \emph{Man-in-the-Middle (MITM) Attack}. During intra/inter-twin interactions, this attack occurs. MITM is an active eavesdropping attack, where adversaries can secretly insert themselves between the two connected entities (e.g., twins or PEs) and possibly alter the communication between them. The attacker may control the entire conversation between two victim nodes, relay messages to them, and make the victim nodes believe that they are directly communicated.
  \item \emph{Sybil Attack}. During inter-twin interactions, Sybil attackers can exploit a single node to simultaneously manipulate multiple active Sybil identities in the decentralized IoDT network with P2P connections \cite{8678450,9852383}. By gaining the majority of influence in the IoDT, Sybil attackers can undermine the power or authority in reputable systems such as 51\% attack in the Bitcoin network.
  \item \emph{Denial of Service (DoS)}. In inter/intra-twin interactions in IoDT, hackers can result a DoS by exhausting the available resources of constrained IoDT devices in the real world. As a consequence, the operations (e.g., simulation and prediction) of digital twins in the digital world can be interrupted. The DoS attack can be caused by the jamming in TCP/IP stack, on-the-path attacks (e.g., blackhole, sinkhole, wormhole, and flooding) at the network layer, or malware injection at the application layer \cite{9765576}. A distributed DoS (DDoS) can be coordinated by compromising multiple nodes to provoke an army of IoDT botnets (e.g., the Mirai).
\end{itemize}

\subsection{Privacy Threats to IoDT}\label{subsec:threat4}

\begin{itemize}
  \item \emph{Pervasive Personal Data Collection}. In intra-twin interaction, to create and evolve an accurate digital clone of the PEs, myriad personal data need to be collected in the IoDT at an unprecedented granularity level and high synchronization frequency, which opens new chances for crimes and misuses of private and sensitive digital twin data.
  \item \emph{Private Information Extraction with Insiders}. This attack occurs in both intra/inter-twin interactions. Insiders can leverage their privileges in the system and its resources to extract security-critical information (e.g., credentials) shared with the digital twin from legitimate end devices or servers. By using this information, attackers can illegally access the digital twins, steal the user's stored personal information, and even carry out cyber espionage. Moreover, after gaining access to the sensitive information, it facilitates potential APT attacks by hackers, ranging from lateral movements within the infrastructure and stealthy manipulations in offering digital twin services.
  \item \emph{Regulation Compliance in Digital Twin Services}. During intra/inter-twin interactions, this attack occurs. To be compliant with privacy regulations like GDPR, authorized service providers should also have user's grant and protect user privacy when collecting/storing/transmitting/processing personal data for big data analysis in offering digital twin services \cite{9268472}.
  \item \emph{Privacy Leakage in Model Aggregation}. There exist potential risks of privacy leakage during digital twin model aggregation process under the collaborative learning paradigm \cite{9928220}. Particularly, the semi-honest cloud/edge server can restore the original training samples through advanced techniques such as the generative adversarial network (GAN) by collecting information such as plaintext gradients, resulting in a risk of data leakage.
  \item \emph{Privacy Leakage in Model Delivery/Deployment}. There exist potential model theft risks in storing and delivering the trained global AI models from the cloud/edge server to participating entities during inter-twin cooperation. If the AI model is stolen, the rich privacy information contained in the AI model parameters may be inferred by the model thief \cite{9645219}. Besides, in the deployment stage of digital twin models, attackers may tamper with the model and implant backdoors, e.g., carefully modifying some neurons. As such, the model behaves normally under normal conditions, but once the backdoor trigger is triggered, the digital twin model's output will be the one preset by the attacker.
  \item \emph{Membership Inference Attack}. This attack exists in both intra/inter twin cooperation. In IoDT, the trained AI models generally no longer rely on the training samples and can map new examples to value predictions or categories via the tuned parameters. However, the process of turning training samples into the AI model is not one way. Via membership inference attacks, adversaries can still inference the sensitive data samples used to train AI models from the model outputs without gaining access to the model parameters \cite{hu2022membership}. Thereby, it results severe model security and user privacy risks for digital twin models trained on sensitive user information.
  \item \emph{Knowledge/Model Inversion Attack}. During the life-cycle of digital twin service, attackers may also extract the representations of the training data in the AI model, known as knowledge/model inversion attacks. Malicious participants may attempt to reveal the private dataset for AI model training by reconstructing each of the classes in the private dataset. The sensitive information extraction from AI models has two types \cite{tramer2016stealing}: (i) directly access the target AI model together with all model structural information (i.e., white-box attack); and (ii) download the target AI model via open APIs and only have model-related information after feeding data to the model (i.e., black-box attack).
  \item \emph{Data Misuse \& Accountability}. In digital twin services, personal and sensitive data can be unintentionally disclosed by authorized service providers or illegally sold out by adversaries for monetary benefits, resulting in huge data misuse concerns. Additionally, due to the easy-to-copy attribute and complex digital twin service cycle, it is hard to trace the misbehaving entities and quickly enforce accountability.
\end{itemize}

\subsection{Trust Issues in IoDT}\label{subsec:threat5}
\begin{itemize}
  \item \emph{Data Trustworthiness}. This threat occurs in both intra/inter-twin interactions. On one hand, as virtual twins are generally untrusted parties without sufficient prior interactions, it raises severe data trustworthiness concerns for data exchange between twins. For instance, the malicious digital twin may share falsified information to mislead others. On the other hand, the synchronized data in real time between PEs and twins can be modified or replaced by adversaries.
  \item \emph{Transaction Fraud}. There also exist inherent transaction frauds in inter-twin data exchanges, resulting in trust and fairness issues \cite{9631953,9632411}. For instance, the seller may sell falsified digital twin models or services and the buyer may refuse to pay at the end of the transaction.
  \item \emph{Free-Riding Threat}. In the open and untrusted IoDT, free-riding PEs or digital twins may behave selfishly to only enjoy the digital twin service without contributing to it \cite{9664267}. For instance, vehicle twins may share redundant information to save the cost of collaboratively training a globally shared AI model for vehicles' route planning.
  \item \emph{Opaque Resource/Knowledge Trading}. Heterogeneous PEs and twins involved in a common task need to collaboratively share their resources or knowledge to improve the efficiency of digital twin services. Besides, a public IoDT market can be created to facilitate resource/knowledge trading. If the resource/knowledge trading behaviors are not transparent, disputes can arise in terms of the resource price, service quality, etc \cite{9632411}. %9632356
\end{itemize}

\subsection{Monetization Issues in IoDT}\label{subsec:threat6}
\begin{itemize}
  \item \emph{Ownership Provenance of Digital Assets}. Compared with physical assets, digital assets can be easily copied and delivered across various platforms, making the ownership provenance of digital assets in IoDT a challenging issue. Moreover, there exist multiple ownership forms (e.g., singly owned or collectively owned) and complex relations between ownership and use right in IoDT, which adds additionally complexity to prove the origin or provenance of digital assets \cite{batista2021blockchains,9631953}.
  \item \emph{Threats to Model Intellectual Property Protection}. The valuable digital twin models can also be stolen for profits via explicit model resell misbehaviors or implicit model extraction behaviors (e.g., model pruning and distillation) \cite{9645219}. The infringement of intellectual property of digital twin models becomes a non-negligible potential threat to the practical deployment of digital twin services.
\end{itemize}

\subsection{Cyber-Physical Threats in IoDT}\label{subsec:threat7}
As the IoDT bridges both the cyber and physical spaces, the IoDT faces two lines of attack: \emph{cyber} and \emph{physical}.
\begin{itemize}
  \item \emph{Physical Damage}. When the digital twin of physical industrial control system (ICS) is compromised, adversaries can learn about the ICS's configuration and illegally access the critical resource via the digital twin to damage the ICS system or exfiltrate critical information. Besides, cyber attacks on critical data of infrastructures can cause damage to physical processes, intellectual property, and control missions.
  \item \emph{Single Point of Failure (SPoF)}. An attacker can launch a physical attack to cause a SPoF of the system due to the destruction of devices/servers, thereby affecting the normal operations (e.g., optimization and monitoring) of digital twin services in the cyber space \cite{9631953}.
\end{itemize}

\subsection{Summary and Lessons Learned}\label{subsec:summary1}
\begin{figure*}[!t]\setlength{\abovecaptionskip}{-0.0cm}
\centering
  \includegraphics[width=19.2cm]{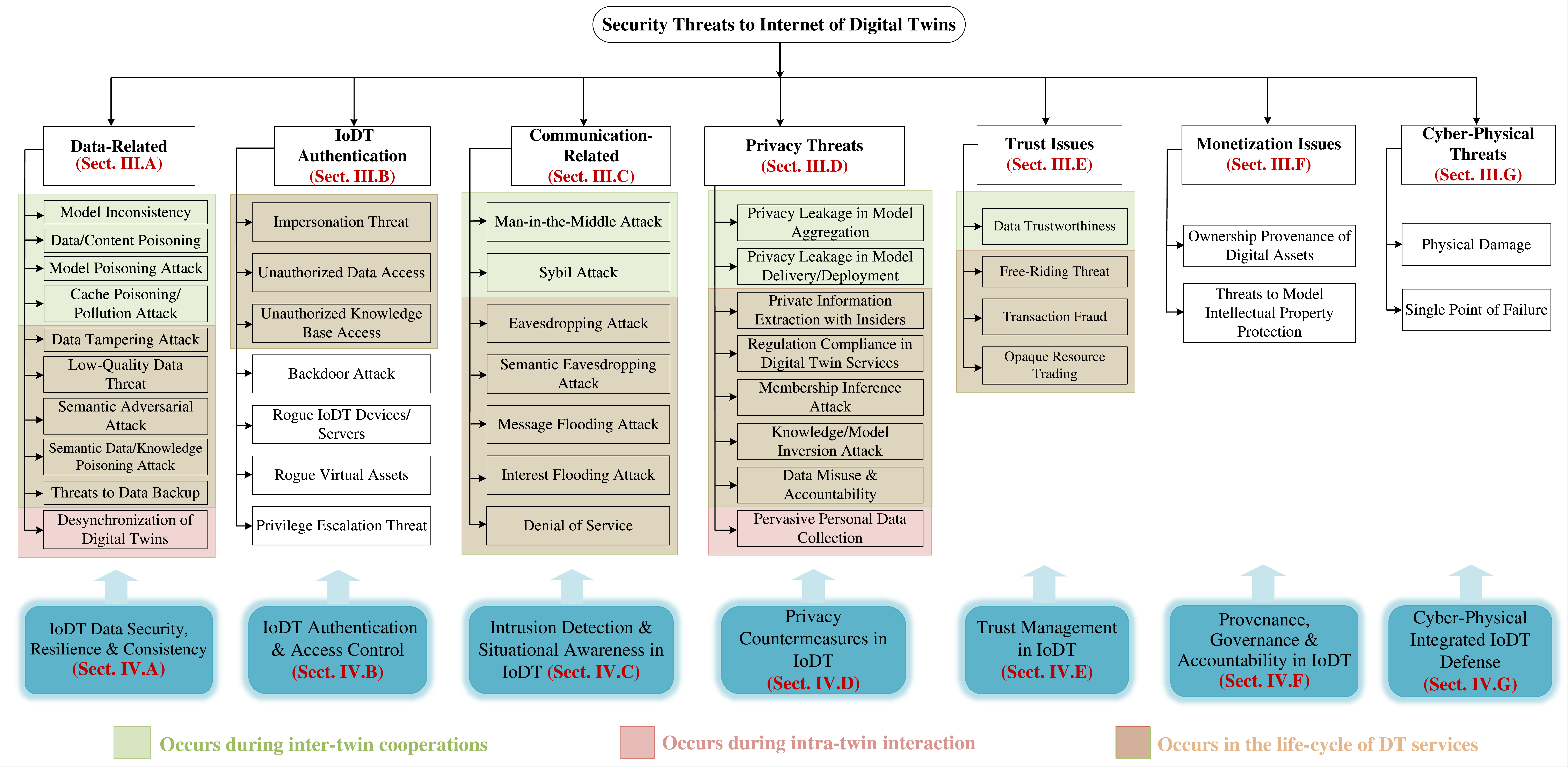}
  \caption{{The taxonomy of security threats to IoDT from seven aspects (i.e., data, authentication, communication, privacy, trust, monetization, and cyber-physical) and corresponding security defenses in the IoDT.}}\label{fig:threatstaxonomy}\vspace{-2mm}
\end{figure*}

As the IoDT is built based on the composition of various cutting-edge technologies, all the existing vulnerabilities, security threats, and flaws can be inherited by the IoDT. Moreover, driven by their interweaving effects and the new features of the IoDT, the impact of existing security/privacy issues in these technologies can be strengthened and become more severe in the IoDT.
Besides, with the increasing diversity and complexity of IoDT functionalities and services, the new IoDT ecosystem can open up opportunities for unexpected threats such as semantic data/knowledge poisoning and breed new types of crimes with more covert, hard-to-trace, and cyber-physical synthesized features.
Lastly, since the IoDT connects both digital and real spaces and requires real-time data feed and feedback between them, it also raises the virtuality-reality synthesized threats such as invasion of state-critical infrastructures via cyber vulnerabilities, as well as the necessities for situational awareness and digital governance.

In the previous subsections (i.e., from Sect.~\ref{subsec:threat1} to Sect.~\ref{subsec:threat7}), we have presented a series of security threats in the IoDT from seven perspectives: data, authentication, communication, privacy, trust, monetization, and cyber-physical. Fig.~\ref{fig:threatstaxonomy} depicts a taxonomy of security/privacy threats in the IoDT. In the next section, we will discuss the state-of-the-art security and privacy countermeasures for IoDT from the above seven aspects in detail.

\section{Security and Privacy Countermeasures in IoDT}\label{sec:Defense}

\subsection{IoDT Data Security, Resilience \& Consistency}\label{subsec:defense1}
\emph{1) Multi-Source Data Fusion in IoDT}.
In the digital twin paradigm, keeping the digital space synchronized with the real space is a basic prerequisite, as any variation between the two spaces can entail significant deviations to the final representation of physical entities/assets \cite{AP238Standard}.
In IoDT, real-time heterogeneous multi-source data fusion is essential to the creation and consistency of digital twins.
To ensure the consistency in autonomous digital twin synchronization, Li \emph{et al}. \cite{9686721} propose a provable data possession method for verifying time states and checking data integrity in virtual spaces. A consortium blockchain ledger is leveraged as the synchronization platform to maintain trusted time state values among distributed physical/virtual entities in IoDT. In their blockchain system, tag verification method is used to prevent legitimate virtual spaces from being framed, and anonymous services are offered to entities for privacy considerations. The work in \cite{9686721} satisfies provable security, conditional anonymity, and unforgeability using rigorous security analysis under the assumption of RSA.

\emph{2) IoDT Data Consistency under Dynamic Constraints}.
The construction of high-fidelity digital twin models is usually constrained by realistic energy supply and data collection strategy. A sustainable data collection method is designed by Wang \emph{et al}. in \cite{9718531} to efficiently build digital twins. To tradeoff the long-term data collection and information loss, a joint optimization method for optimizing both reveal delay and data fidelity under constraints for sustainable information and energy is also developed in \cite{9718531}. Both analytical and simulation analysis demonstrate the feasibility of their method.
In addition, the estimation and analysis of real-time environmental and structural factors in dynamic synchronization between the PE and its virtual representation are challenging issues, especially for multiple small objects in complex and large-scale scenes.
To address these issues, Zhou \emph{et al}. \cite{9363597} consider the equipment, operator, and product as the basic factors to analyze the dynamics in constructing a generic digital twin system. Based on feature fusion from both deep and shallow layers, a learning-based algorithm is also devised for efficient detection of multi-type small objects. Thereby, the modeling, monitoring, and optimization of physical manufacturing processes can be facilitated with the aid of virtual twins.

Gehrmann \emph{et al}. \cite{8822494} identify the security issues of digital twins in terms of synchronization, software, network isolation, and DoS resilience. Moreover, a novel security architecture based on the Dolev–Yao model is presented, and a new state replication and synchronization mechanism is designed to satisfy expected synchronization requirements of digital twins. A proof-of-concept (PoC) implementation using programmable logic controllers (PLCs) is presented to assess the proposed design's components and security performance.

\emph{3) Blockchain for IoDT Data Security}.
In IoDT, conventional cloud/fog-enabled twins usually suffer from typical sensitive information leakages, data manipulation, and data reliability issues due to the malfunction of cloud/fog servers.
To resolve the above issues, Khan \emph{et al}. \cite{9310354} propose a novel blockchain-based spiral framework of digital twins, where a new blockchain variant
called \emph{twinchain} is devised to resist quantum attacks and provide instant transaction confirmation. A case study on the manufacturing of a surgical robot validates the proposed twinchain's applicability.
To further reduce the operation cost and secure digital twin-related transactions, Liao \emph{et al}. \cite{9660773} deploy a permissioned blockchain and auction-based pricing mechanism for dynamic service matching in intelligent transportation system (ITS) between digital twin service providers and requesters. To improve consensus efficiency, a novel DT-DPoS (digital twin delegated proof of stake) consensus protocol is also designed to better suit the digital twin-enabled ITS scenarios.

Several research efforts have been reported in the literature to secure and optimize digital twin-based applications such as industrial metaverse \cite{9880528,hashash2022towards}, vehicular traffic management \cite{9875052}, maritime transportation systems \cite{9626558}, industrial IoT \cite{9832511}, edge offloading \cite{9885226,9174795,9166745}, and virtual reality (VR) \cite{9964320}.

\emph{4) IoDT Data Synchronization in Metaverse}.
Digital twin is a supporting technology for the industrial metaverse, and the seamless synchronization of distributed digital twins and their associated sub-metaverses at the wireless edge is essential to build a decentralized metaverse framework. In \cite{hashash2022towards}, Hashash \emph{et al}. design an IoDT system comprised of autonomous cyber twins and physical twins operating in massively-sensed edge environment, where a problem with optimization is formulated to minimize the sub-synchronization latency between digital and physical spaces while satisfying synchronization intensity requirements of cyber twins. The optimal transport theory is employed for problem solving as well as allocation of computation and communication resources. Simulation results show a 25.75\% reduction in sub-synchronization delay between cyber twins and sub-metaverses.

\emph{5) IoDT Data Synchronization in ITS}.
For traffic management in vehicular ad hoc networks (VANETs), the use of digital twin can help map the traffic conditions on the real road environments into the cyber world. However, there exist potential data security and reliability issues in digital twin-enabled vehicular traffic management tasks. Feng \emph{et al}. \cite{9875052} design a vehicular blockchain to construct a decentralized virtual twin model for the in-vehicle self-organized network with satisfactory performance (i.e., communication overhead less than 700 bytes, stable message delivery rate at 80\%, and data leakage rate at about 10\%).
Liu \emph{et al}. \cite{9626558} focus on the data relay security in digital twin-enabled collaborative maritime transportation systems, and propose an optimization scheme for maximized secrecy rate with low transmission delay in the maritime communications.

\emph{6) IoDT Data Synchronization in Industrial IoT}.
Digital twin-enabled industrial IoT usually relies on cloud/edge servers for compute-intensive and real-time data processing. Aimed to mitigate the unreliable public communication channels and build trust among participating entities, Kumar \emph{et al}. \cite{9832511} integrate deep learning and blockchain to deliver decentralized data learning and digital twin services in industrial IoT. The smart contracts are deployed atop the blockchain platform to guarantee data integrity and authenticity, and an intrusion detection system (IDS) is built based on long short term memory (LSTM), sparse autoencoding (SAE), and multi-head self-attention (MHSA) techniques to make sure the information obtained from the blockchain is accurate. Evaluations on the implementation of the proposed framework demonstrate a significant improvement in data privacy and communication security.

\emph{7) Edge Offloading in IoDT}.
To alleviate the intensive computation in digital twin creation and update, computation offloading is a promising approach. Digital twins can help offload decisions in wireless edge networks, where digital twins corresponding to edge nodes estimate the states (e.g., computation capacity) of edge nodes to optimize offloading decisions.
Huynh \emph{et al}. \cite{9885226} leverage digital twins to model edge nodes' computation capacity and optimize resource allocation in terms of edge processing latency, transmission latency, and local processing latency. An alternating optimization method and inner convex approximation method are also studied to solve the formulated problem with non-convex constraints in an iterative manner.

Sun \emph{et al}. \cite{9174795} study a mobile offloading scheme in digital twin edge networks (DTENs) to reduce offloading latency while accounting for user mobility and service migration costs. A Lyapunov optimization approach is developed to simplify the constraint, and an actor-critic RL method is proposed to solve the optimization problem. Simulations show that their scheme with digital twins outperforms existing works in reduced offloading latency, service migration rate, and offloading failure rate. Considering the resource-limited IoT devices, resource heterogeneity and stochastic tasks in DTENs, Dai \emph{et al}. \cite{9166745} further leverage Lyapunov optimization and asynchronous actor-critic algorithm to derive the optimal stochastic offloading strategy in digital twin-enabled edge networks.

\emph{8) IoDT QoS Optimization in VR Systems}.
By integrating VR and digital twin technologies, VR-embedded digital twins (VR-DT) can facilitate the visualization of digital representations of manufacturing in the industrial IoT.
Concerning the data-driven, security-sensitive, and compute-intensive features, Song \emph{et al}. \cite{9964320} offer a blockchain-based decentralized resource allocation framework in VR-DT service offering under industrial IoT with reduced service latency and improved transaction throughput. A mixed-integer nonlinear programming (MINLP) problem is formulated to jointly optimize the QoS in VR-DT in terms of channel allocation, computation capacity assignment, subframe configuration, and block size adjustment. A multi-agent compound-action actor-critic algorithm with full decentralization is also devised to resolve the QoS optimization issue. Experimental results demonstrate the superiority of the proposed framework in enhancing the QoS of VR-DT services, in comparison with existing benchmarks.

\emph{9) IoDT Data Resilience}.
For enhanced data resilience in harsh environmental areas such as disasters and mountains, existing works on air-ground collaborative networking \cite{9696188} and robust blockchain design \cite{9488719} can offer some lessons for the provisioning resilient and efficient digital twin services. %,wang2022secure  9035635,

\subsection{IoDT Authentication \& Access Control}\label{subsec:defense2}
\emph{1) IoDT Authentication in IoV}.
As a typical IoT scenario, there are increasing works on the IoDT authentication under vehicular environments.
In the cloud-based Internet of vehicles (IoV), Xu \emph{et al}. \cite{9625518} propose two novel authentication protocols for both intra-twin and inter-twin communications based on the group signature and secret-handshake scheme. Strict security analysis proves the conditional anonymity and unlinkability of physical/virtual vehicles.
By further considering vehicle mobility in edge-enabled IoV, Li \emph{et al}. \cite{9645213} design a security reference architecture for digital twin-driven IoV and devise a handover authentication method based on proxy ring signatures to realize cybertwin migration and mutual authentication between on-road vehicles and the road-side edge node. Simulations on a computer using the OpenSSL tool show the efficiency of the proposed architecture in respect of computation overhead and bandwidth consumption.

\emph{2) Blockchain for IoDT Authentication in IoV}.
In \cite{9810813}, the blockchain is further employed by Liu \emph{et al}. to prevent impersonation and assist IoDT authentication, where a group authentication method with privacy preservation is proposed in digital twin-enabled IoV to mitigate impersonation threats. In \cite{9810813}, nodes' public keys are stored in the public blockchain ledgers to ensure transparency, and a GAN-based method is devised for risk forecast of twins in IoV. Simulation results validate the proposed IoV group authentication method outperforms conventional ones in terms of defensive performance.

\emph{3) AI and Blockchain for IoDT Authentication in Smart Grid}.
Apart from the IoV, some works have explored the IoDT access control scheme in smart grids. For instance, Lopez \emph{et al}. \cite{9430900} develop an AI and blockchain enabled intelligent authorization method in smart grids, where the AI-based semantic platform enables feature prediction and optimization while the blockchain-based authorization platform enforces automatic access control. Based on the transparent blockchain ledgers, the access policy decision points in local domains can be coordinated to reach consensus on the global access policy decisions.

\emph{4) Access and Usage Control in IoDT}.
To implement access control policies, the attribute-based encryption (ABE) schemes including key-policy ABE (KP-ABE) and cipertext-policy ABE (CP-ABE) can be employed depending on the specific applications.
Additionally, smart contracts can be utilized to enable automatic and fine-grained access control in the IoDT. For instance, the SPDS \cite{9268472} utilizes the smart contracts on top of the blockchain to stipulate fine-grained data access and usage policies in aspects of who can access what types of data, under what conditions, and for what purposes. For privacy concerns in public smart contract environments, there have been growing interests in combining smart contract and trusted computing technologies \cite{8806762,9582795,9445602,9999528}. For instance, in \cite{9268472}, a trust processor is utilized to process confidential user data in an off-chain manner and record data usage activities on distributed ledgers in an immutable manner. For efficient coordination of on-chain and off-chain contract execution, an atomic delivery protocol with two phases is also devised in \cite{9268472} to ensure the transactional atomicity.
Besides, to ensure privacy preservation of digital twins and PEs in the smart contracts, existing researches on advanced cryptographic tools such as homomorphic encryption (HE) \cite{9833732} and zero knowledge proof (ZKP) \cite{7961949} can offer some lessons.

\subsection{Intrusion Detection \& Situational Awareness in IoDT}\label{subsec:defense3}
\emph{1) Intrusion Detection of IoDT in ICS}.
IoDT, as a rising digital system combining physical-cyber interactions, makes it more convenient to detect intrusions and anomalies in CPS in a timely and accurate manner. To guarantee the stability and efficiency of IoDT systems, there have been various works on intrusion detection in IoDT.
To resist cyber threats for ICS, Li \emph{et al.} \cite{9999272} present a terminal-to-terminal detection mechanism to realize real-time and accurate anomaly detection. To facilitate subsequent feature extraction, the multidimensional deconvolution approach is adopted to obtain the low-dimensional characteristics of the original data from the input of high-dimension. Extensive simulation results demonstrate the advantages on detection precision in comparison with benchmark methods. Taking into account the complex industrial environments and network heterogeneity, Bellavista \emph{et al.} \cite{bellavista2021application} exploit an application-enabled digital twin system to simplify the management of network resources.

\emph{2) Intrusion Detection of IoDT in ITS}.
Accurate traffic streaming prediction and intrusion detection are crucial issues in ITS. The IoDT-enabled secure ITS has been studied in works \cite{9560727,9645156}. In \cite{9560727}, the deep learning-based method is proposed to secure digital twin-enabled cooperative ITS, in which data characteristics of traffic congestion generated from emergencies are used to train the traffic digital twin model for online real-time prediction. In \cite{9645156}, Yin \emph{et al.} propose a cybertwin-enabled secure transmission scheme in satellite-terrestrial integrated vehicular networks, where the global information sharing and cooperation between satellite and terrestrial networks are implemented in cybertwins.

\emph{3) Situational-Aware IoDT}.
The success of IoDT also requires efficient situational awareness of data sources to track the accountable entity for creating or updating digital twins. Several studies have investigated situational awareness approaches to safeguard IoDT-based frameworks.
To support situational-awareness environments, Suhail \emph{et al.} \cite{suhail2022towards} present a blockchain and digital twin framework as trusted twin towards situation-aware CPS. To ensure reliable system data, the data sources truthfulness via integrity checking mechanisms (ICMs) is deigned in \cite{suhail2022towards} to model the process knowledge of digital twins.
The digital twin for situational awareness in industrial systems is investigated in \cite{varghese2022digital} for malicious attacks and defense simulation, in which four types of process-aware attack scenarios (i.e., command injection, DoS, and naive/computed measurement modification) are exploited. Simulations validate the advantages of the designed stacked model for real-time intrusion detection. Considering the autonomous core networks, Yigit \emph{et al.} \cite{9927259} present a digital twin-assisted DDoS detection scheme through an online learning approach. Xiao \emph{et al.} \cite{xiao2022commandfence} investigate a digital twin-based security framework to protect the smart home system. Deep learning is a promising approach for intrusion detection. In \cite{zhang2020cyber}, a new deep neural model of IoDT is proposed for recognizing potential vulnerable functions in smart healthcare. In \cite{9889287}, the storage security of edge-fog-cloud for deep learning-assisted digital twin is proposed to guarantee the storage security.

\emph{4) Placement and Migration of Digital Twins}.
The dynamic network states and environment, such as available computation and communication resources, may limit digital twins from promoting QoS performance. The placement and maintenance of IoDT is a fundamental problem that should be well addressed. By integrating digital twins with edge network, Lu \emph{et al.} \cite{9491087} propose a wireless DTEN model and formulate an edge association problem between edge nodes and digital twins to determine the placement of digital twins in the proposed framework. Numerical results have demonstrated the improved convergence rate in complex network scenarios.

\subsection{Privacy Countermeasures in IoDT}\label{subsec:defense4}%{\color{blue}   }
\emph{1) Blockchain for Privacy Preservation in IoDT}.
Labeling and tracking physical objects are of great significance for various complex systems in IoDT. Since the IoDT requires real-time data acquisition from physical systems, the privacy of digital twins and physical systems/entities should be well-protected. There have been various works on privacy preservation in IoDT via blockchain approaches \cite{lu2020communication,9606227,son2022design}. Lu \emph{et al.} \cite{lu2020communication} utilize the DTEN to guarantee the synchronization for the integration of edge networks and digital twins. To protect data privacy, the blockchain-integrated federated learning scheme is also presented to ensure data privacy protection. Theoretical analysis validates communication efficiency and data security. Jiang \emph{et al.} \cite{9606227} study a DTEN framework to implement a flexible and secure digital twin platform, where federated learning is exploited to establish the IoDT model. In order to guarantee the security of local model and global model updates, a blockchain platform for model updates is also designed. Son \emph{et al.} \cite{son2022design} design a privacy-preservation scheme to secure IoDT data sharing and communication in cloud-enabled digital twin networks. The cloud computing is exploited for facilitating data sharing, and the blockchain is adopted for data verifiability and privacy preservation in IoDT.

\emph{2) Federated Learning for Privacy Preservation in IoDT}.
Federated learning, as a distributed AI paradigm, allows clients to train machine learning models locally without uploading local private data to the cloud. Federated learning is a promising technique to attain a trade-off between user privacy protection and the utilization of decentralized big data for constructing IoDT models. Researchers have investigated the integration of IoDT and federated learning \cite{chen2021digital,9311405,9170905}. Chen \emph{et al.} \cite{chen2021digital} investigate the edge-empowered and digital twin-based distribution estimation federated learning scheme. In federated analytics, the personal data is not shared within digital twins, which protects the users' privacy. Numerical results demonstrate the accuracy and convergence of the federated analytics compared with benchmark schemes. Sun \emph{et al.} \cite{9311405} propose an incentive-enabled dynamic digital twin and federated learning framework, where wirless devices train the local models using their local data instead of transmitting the natural data to servers to guarantee data privacy. Taking varying digital twin deviations into account, the incentive mechanism is provided to select the optimal clients for participation. Numerical results validate the effectiveness and efficiency of the designed framework in improving model accuracy. By migrating the digital twins into wireless communication networks, Lu \emph{et al.} \cite{9170905} exploit the digital twin wireless networks (DTWNs) to improve the efficiency of data processing. The designed blockchain and federated learning are operated in the proposed DTWN to guarantee the reliability of DTWNs while ensuring data privacy protection for users. Numerical results testing on real-world datasets have validated the performance advantages of DTWN.

IoDT can provide guidance for multidimensional resource allocation via building a digital representation of the physical entities. Zhou \emph{et al.} \cite{9658206} design a federated learning-enabled digital twin framework and propose a digital twin-based resource scheduling algorithm to guarantee the digital twin system with low-latency, accurate, and secure performance. Simulation results show that SAINT has superior performance in comparison with state-of-the-art algorithms. Schwartz \emph{et al.} \cite{schwartz2021linking} propose a typical markers for invisibility to users via IoDT. Through adding artificial markers to indoor and outdoor, the mapping of the scenarios is advocated to provide reliable and secure information to robots, with the objective of enhancing the reliability of robotic navigation and decreasing computational cost.

\emph{3) Other Technologies to be Explored}. Apart from blockchain and federated learning technologies, other privacy computing technologies including differential privacy (DP), secure multi-party computing (SMC), and HE can provide some lessons for privacy protection in the life-cycle of digital twin services in IoDT.

\subsection{Trust Management in IoDT}\label{subsec:defense5}
\emph{1) Trust Evaluation and Trust-Free Approaches}.
IoDT depends on trustworthy sensory/processing data from the physical/cyber worlds for reliable decision-making and feedback. As such, IoDT should be able to make reliable decisions through identifying faults based on these uncalibrated data. High-fidelity is one of the key challenges for creating virtual model in IoDT. The trust management plays an important role in IoDT to ensure the data trustworthiness for building high-fidelity digital twins. Representative researches in this context can be classified into two lines, i.e., quantitative trust evaluation approaches \cite{9834331,8740953,6081879} and blockchain-based trust-free approaches \cite{9357935,9362182,9001017,9833297,9954886,botta2021secure}. %9048619,
For trust evaluations, Wang \emph{et al.} \cite{8740953} design a quantitative trust model by integrating the direct and indirect trust evaluations. Das \emph{et al.} \cite{6081879} develop a dynamic trust model by considering the recent trust, historical trust, expected trust, and trust decay for global trust computation.
Blockchain, as a decentralized ledger, provides a promising solution with salient features including trust, accountability, data integrity, and immutability to assist trust-free interactions in IoDT.
For trust-free digital twin creation, Suhail \emph{et al.} \cite{9357935} present a blockchain-based mechanism to deal with the issues of data management and security in digital twins, thereby guaranteeing the trustworthiness of data sources.
Raes \emph{et al.} \cite{9362182} further propose a novel framework to construct interconnections and reliable digital twins in smart cities. The proposed digital twin models can timely interact with the smart city in diverse domains (e.g., transportation, environment, and health) from different data sources.

\emph{2) Blockchain for Trust Management in IoDT Services}.
In IoDT, the data records of collaboration activities between different virtual twins should be reliably documented to ensure traceability and trust. There have been several studies exploiting blockchain for trust management in IoDT data management. Hasan \emph{et al.} \cite{9001017} present a blockchain-based digital twin creation scheme to ensure trusted traceability and data provenance via smart contracts. The decentralized storage system is used to store and share digital twin data.
Test results show that the proposed approach satisfies the requirements of digital twin process creation. Gai \emph{et al.} \cite{9833297} design a blockchain-based digital twin framework to support chain management (SCM) system, in which the blockchain is adopted for trusted data storage and tracing in digital twin implementation. Experiments demonstrate the efficiency and effectiveness of the digital twin-based SCM system. By integrating blockchain and digital twins, Zhang \emph{et al.} \cite{9954886} propose a blockchain and digital twin-empowered smart parking system. The digital twin system is utilized to monitor and analyze traffic conditions in real-time, and the blockchain platform is used to manage trust values and offer reliable data storage. To enhance the robustness of trust management system, the blockchain-based supply chain management is also proposed in \cite{botta2021secure} for verifiable digital twins, in which each PE has an identified digital twin linked by a unique code in the blockchain.

\emph{3) Trust-Based Model Aggregation in IoDT Services}.
Apart from the blockchain technology for trust management, several works have investigated the trust-based aggregation for federated learning. Qu \emph{et al.} \cite{9839634} provide an asynchronous federated learning (FedTwin) scheme to guarantee privacy-preservation in IoDT via blockchain. In local training stage, the GAN-empowered differential privacy is defined to protect the privacy in local model parameters by adding the noise. In global model aggregation, an improved Markov decision approach is utilized to determine the optimal digital twin for asynchronous aggregation. Sun \emph{et al.} \cite{9244624} design a novel architecture of digital twin-empowered IoT and propose an adaptive federated learning framework. To enhance the reliability and accuracy of learning models, clients' contribution  to the global aggregation is quantified by measuring the deviation of digital twin from the trust-weighted aggregation strategy. Dai \emph{et al.} \cite{9773095} investigate a digital twin-envisioned secure federated aerial learning framework. To ensure trustworthy federated learning models, the blockchain ledgers are utilized to guarantee the security in data transmissions under federated learning.

\subsection{Provenance, Governance \& Accountability in IoDT}\label{subsec:defense6}
\emph{1) Blockchain for IoDT Provenance and Governance}.
Traditional cloud-based centralized architecture for digital twin service offering usually lacks flexibility and is prone to SPoF risks. Various works \cite{9129796,9384115,9750989} have exploited the promising blockchain technology to build a decentralized and flexible digital twin realm.
Concerning the poor flexibility and SPoF issues under the cloud-based centralized architecture, Zhang \emph{et al}. \cite{9129796} leverage the permissioned blockchain technology to design a manufacturing blockchain architecture in the digital twin manufacturing cell. In their architecture, both hardware equipment and software-defined components are developed to improve manufacturing efficiency. A prototype of the proposed architecture is designed, and the evaluation experiment show its satisfactory throughput and latency performance.

To further enforce auditability and traceability of critical data, Wang \emph{et al}. \cite{9384115} investigate a two-layer blockchain-based framework in the hemp supply chain and design a digital twin model based on stochastic simulation for risk management with dynamic evolution and spatial-temporal causal interdependencies. In the proposed blockchain, state regulators and local authorities can run the proof-of-authority (PoA) consensus protocol to enforce transparent quality control verification.
To resolve security issues in knowledge trading while ensuing high reliability and low latency, Wang \emph{et al}. \cite{9750989} investigate a novel blockchain-empowered hierarchical digital twin framework in edge-enabled IoT context. A dual-driven learning approach for both data and knowledge is designed to enable real-time interaction between physical and cyber spaces. Moreover, a proximal policy optimization (PPO) method is devised in the multi-agent RL process to minimize energy consumption and overall latency. Numerical results show that the proposed approach can improve learning accuracy, enhance system reliability, and balance energy consumption and system latency.

\emph{2) Deep Learning for IoDT Governance}.
Deep learning technologies can assist deliver secure and regulatable digital twin services.
Lv \emph{et al}. \cite{9560727} combine deep learning and digital twin technologies for enhanced road safety in the ITS. Both convolutional neural network (CNN) and support vector regression are involved for improving prediction accuracy. The simulation results show that their proposed approach achieves a high security prediction accuracy of 90.43\% to reduce the effect of traffic congestions.

\emph{3) Game-Theoretical IoDT Governance}.
Apart from the solutions built on blockchain and AI technologies, game-theoretical approaches have been widely investigated in the literature for attack defense \cite{9095990}, service congestion governance \cite{9800199}, and long-term incentive design \cite{9583902}.
Xu \emph{et al}. \cite{9095990} identify a novel \emph{stealthy estimation threat}, where smart attackers can learn defense strategies to alter the digital twins' state estimation without being detected. To produce the online digital model corresponding to the real-world system, a Chi-square detector is designed. In addition, to seek the optimal attack and defense policies, a signaling game approach is investigated. The proposed game theoretical approach can lessen the attack impact on the PEs and enforce the stability of the CPS, according to both analytical and experimental results.

\emph{4) Incentive Design for IoDT Governance}.
In IoDT, the intensive and dynamic virtual twin service demands can easily result in service congestion, which eventually deteriorates the QoS and stability of digital twin services.
Peng \emph{et al}. \cite{9800199} study a digital twin-empowered two-stage offloading mechanism in DTENs for mitigating latency-critical tasks from end devices to edge servers. In the first stage, credit-based incentives are assigned to optimize digital twins' resource allocation strategies; while in the second stage, a Stackelberg game is designed to derive the optimal offloading and privacy investment policies for digital twins. Experimental results show that the proposed mechanism realizes efficient computation offloading while guaranteeing data privacy.

Considering the spatio-temporal dynamic demands of digital twin services, Lin \emph{et al}. \cite{9583902} investigate the DTEN's long-term effective incentive-driven congestion control scheme. The long-term congestion control problem is decomposed into multiple online edge association subproblems with no future system information dependencies using Lyapunov optimization method. A contract-theoretical incentive mechanism is devised to maximize the digital twin service provider's utility, with consideration of individual rationality (IR), incentive compatibility (IC), and delay sensitivity. Using the base station dataset of Shanghai Telecom, simulation results show the efficiency of their proposed scheme in long-term service congestion mitigation compared with benchmarks.

\subsection{Cyber-Physical Integrated IoDT Defense}\label{subsec:defense7}
\emph{1) Digital Twin for Protecting Physical Systems/Infrastructures}.
The emerging digital twin technology is promised to mitigate the increasing cyber-attacks on physical systems such as ICS \cite{8966454} and critical infrastructures such as power grids \cite{9368968,9112234,9000371}, as well as ensure public safety \cite{9267778} and alleviate COVID-19 pandemic \cite{9409764}.
For instance, Saad \emph{et al}. \cite{9112234} deploy digital twins in the IoT cloud to improve the resiliency of interconnected microgrids and promote the digital twin-as-a-service (DTaaS) paradigm. In their work, digital twins can interact with the physical control system (which is implemented by single-board computers) to resist DoS and false data injection attacks and enforce proper system operations. Real implementations on Raspberry and remote AWS cloud show the feasibility and effectiveness of their proposed system in attack defense.
Moreover, Marai \emph{et al}. \cite{9267778} deploy a digital twin box (DTBox) on road infrastructures to produce digital twins of road assets via real-time data transmission (e.g., live stream of camera) to/from the cloud/edge. An object detection module is also designed inside the DTBox to identify and track specific objects including vehicles and persons from the captured live stream to enhance public security.
Besides, in the Elegant project \cite{9499077}, digital twins are created and deployed based on high-fidelity virtual replicas of PLCs to alleviate security risks such as DDoS with the assistance of AI models. Experiments on Fed4Fire federated testbeds validate its feasibility in utilizing digital twins with data pipelines to defend against DDoS attacks.

\emph{2) Digital Twin for Live/Postmortem Forensics}.
Dietz \emph{et al}. \cite{8966454} introduce multiple security-operation modes in ICS enabled by digital twins including replication, historical data analytic, and simulation to facilitate live and postmortem digital forensics.
By operating in the replication mode, digital twin can mirror the current events and states of ICS to detect cyber-attacks. By analyzing digital twin's historical database, the attack time, point of origin, and subsequent lateral movements of stealthy attackers can be detected. Additionally, the malicious activities can be replayed by operating in simulation and replication modes, where the simulation mode replicates various attack versions by learning from the historical database. Thereby, the back-tracing of attack behaviors can be enabled to facilitate live and postmortem forensics.

\emph{3) Economic and Social Effects in Defenses}.
However, existing advanced digital twin services in CPS mainly focus on performance, including accuracy and processing speed, while the economic and social costs are usually ignored.
Aiming for an eco-friendly IoDT instead of a performance-biased one, Kim \emph{et al}. \cite{9932580} propose a green AI-enabled digital twin security surveillance framework with low resource consumption. The optimization problem to motivate the participation of reusable devices for eco-friendly security is expressed as an integer linear programming (ILP) problem, which is solved by the designed dense sub-district method.
Numerical results demonstrate the effectiveness of their proposed framework in terms of resource consumption to ensure a satisfactory surveillance range.

%\subsection{Summary and Lessons Learned}\label{subsec:summary2}

%%Owing to the large amount of data, stringent delay requirements, and computation-intensive 3D simulations, real-time cyber-physical interactions become a challenging issue.

\section{Future Research Directions}\label{sec:FUTUREWORK}
In this section, we discuss several future research directions in the field of IoDT from the following aspects.

\subsection{Cloud-Edge-End Orchestrated IoDT}\label{subsec:future1}
The explosive growth of terminal equipment has led to serious loads in IoDT for processing big data. The end-users may not be served seamlessly by the IoDT system during the service period, which suffers from service interruptions when users move outside the coverage of the access points associated with the twin. The cloud-edge-end orchestrated architecture, which is composed of the cloud tier, edge tier, and end tier, can collaboratively establish the service function chain (SFC) for enhanced QoS \cite{9268472}. The cloud tier has powerful computing capability, which can provide sufficient computing power for AI model training and intelligent analysis. The edge tier is located nearer to the data source, which can facilitate real-time processing and high efficiency in data synchronization \cite{9664267}. The cloud-edge-end orchestrated IoDT architecture can achieve on-demand resource sharing and feasible networking for massive PEs and digital twins. Besides, each twin of the end-user exists in the cloud or edge server, and each twin acts as the agent to improve the quality-of-experience (QoE) for end-users. Future works can be investigated including the dynamic resource collaboration, multi-layer and multi-dimensional resource allocation, and intelligent application systems for the cloud-edge-end orchestrated IoDT.

\subsection{Space-Air-Ground Integrated IoDT}\label{subsec:future2}
Space-air-ground integrated networks (SAGIN) \cite{yin2022cybertwin}, which connect multi-tier networks including the space subnetworks, air subnetworks, and ground subnetworks, hold great potential to meet the QoS needs of 6G networks such as ubiquitous coverage and ultra-wide-area broadband access. In light of the upcoming challenges (e.g., security, privacy, and dynamic network environment) in SAGIN, service performance may be affected by heterogeneous resources and diverse network protocols \cite{9631953}. IoDT has the ability to decrease decision risks and strengthen service intelligence via AI technologies for SAGIN and the virtual space. As such, space-air-ground integrated IoDT provides a promising potential to solve the challenges in complicated network situations, enabling efficient operations and management in SAGIN. Future research directions toward space-air-ground integrated IoDT still include real-time cross-domain authentication, integrated sensing, communication and computing, and collaborative blockchain deployments.

\subsection{Interoperable and Regulatory IoDT}\label{subsec:future3}
The interoperability of IoDT refers to as the capacity of system to freely exchange information across various digital twins in the cyberspace, as well as between physical and cyber spaces \cite{9765576}. The interoperability of the IoDT includes various aspects including hardware, software, protocols, interfaces, and even operating systems, which requires multi-dimensional efforts from both industry and academia. Open research challenges towards interoperable IoDT include the design of all-around new standards and cross-chain interoperable mechanisms.
Moreover, regulations are essential to the future development of the IoDT system to delimit disputes, track/decide criminal behaviors, enable digital forensics, and enforce punishments in the new IoDT ecology. AI and blockchain technologies can empower IoDT governance. For instance, AI can enable misbehavior detection, association of twin-activity, and AI-based judge; while blockchain allows automatic law-enforcement using smart contracts and decentralized and democratic governance via distributed consensus mechanisms.
Open research challenges towards regulatory IoDT include the design of new ``hard law" and ``soft law" \cite{9880528}, explainable AI algorithms, smart contract protection, IoDT-specific consensus mechanisms, and regulated blockchains \cite{9631953}.

\subsection{Explainable AI-Empowered IoDT}\label{subsec:future4}
In IoDT, AI technologies can help produce and evolve digital twins with high fidelity and consistency, enable adaptable semantic communications, establish security situation awareness platforms, and build regulatory IoDT. As such, the explainability of AI-based decisions is of significance to guide the IoDT development and help improve AI algorithms \cite{9233366}.
As an effort, Tripura \emph{et al}. \cite{tripura2022prob} design an interpretable machine learning for digital twin updating by using interpretable physical and mathematical functions to express the dynamics of a real system. Based on sparse Bayesian regression, only the critical parts representing the perturbation terms in the underlying dynamics of physical twins are accurately identified in \cite{tripura2022prob} to update digital twins.
However, future works to be investigated for explainable AI in IoDT still include learning semantics of AI model components and the generation of explanations. %making parts of AI model transparent,

%\subsection{Information Bottleneck Based Privacy-Aware IoDT}\label{subsec:future5}

\section{Conclusions}\label{sec:CONSLUSION}
In this paper, we have presented a comprehensive survey on the working principles, security and privacy, and future prospects of IoDT.
Firstly, a novel distributed IoDT architecture with cyber-physical interactions is introduced, along with the information flows across digital twins and their physical counterparts via inter-twin and intra-twin communications.
Then, the supporting technologies to build an IoDT engine and the critical characteristics of IoDT are discussed.
Furthermore, we have investigated a taxonomy of security and privacy threats in IoDT, as well as the key challenges in security defenses and privacy protection under the distributed IoDT architecture.
We have also reviewed the state-of-the-art security and privacy countermeasures to design tailored defenses approaches in IoDT. Finally, future research directions essential to IoDT are discussed.
The main goal of this survey is to provide a thorough and in-depth understanding of IoDT working principles including its general architecture, key characteristics, security/privacy threats, and existing/potential countermeasures, while inspiring more pioneering efforts in the emerging IoDT paradigm.

\bibliographystyle{ieeetr}

\bibliography{ref.bib}
\end{document}